\documentclass{cimento}
\usepackage{epsf}
\title{On structure of solutions of 1-dimensional 2-body problem\\
in Wheeler-Feynman electrodynamics}
\shorttitle{2-body problem in Wheeler-Feynman electrodynamics}
\author{S.V.Klimenko\from{ins:x}, 
I.N.Nikitin\from{ins:y}, 
W.F.Urazmetov\from{ins:x}}
\instlist{
\inst{ins:x} Institute for High Energy Physics, Protvino, Russia
\inst{ins:y} National Research Center for Information Technology,
St.Augustin, Germany}
\PACSes{
\PACSit{03.20}{classical mechanics of discrete systems}
\PACSit{03.50}{classical field theory}
}

\newlength{\mywidth}\mywidth=6cm 
\def\fref#1{fig.\ref{#1}}
\def\bfref#1{(fig.\ref{#1})}
\renewenvironment{figure}{%
    \refstepcounter{figure}
    \baselineskip=0.4\normalbaselineskip\footnotesize%
}
{\baselineskip=\normalbaselineskip}
\def\fignum{{\bf Fig.\arabic{figure}.\quad}}
\def\figl#1#2#3{%
    \begin{figure}\label{#1}
      \begin{center}
        \vspace{0.2cm}
        #2
        \quad
        \parbox{\mywidth}
        {\fignum #3}
      \end{center}
      \vspace{0.2cm}
    \end{figure}%
}
\newenvironment{quot}
{\begin{quotation}\baselineskip=0.4\normalbaselineskip\footnotesize}
{\baselineskip=\normalbaselineskip\end{quotation}}
\def\ln{\mathop{\rm ln}\nolimits}
\def\ch{\mathop{\rm ch}\nolimits}
\def\sh{\mathop{\rm sh}\nolimits}
\def\min{\mathop{\rm min}\nolimits}
\def\max{\mathop{\rm max}\nolimits}
\def\d{\partial}
\def\R{{\bf R}}
\def\Tau{{\cal T}}
\def\half{{\textstyle{{1}\over{2}}}}
\def\nn{\nonumber}
\def\ortograph#1{\noindent #1}
\def\mar{~$\!\!\!\!\!\!\!\!\!\!\!\!\!\!\!\!\!$~}
\begin{document}

\maketitle

\begin{abstract}
The problem of 1-dimensional ultra-relativistic scattering
of 2 identical charged particles in classical electrodynamics
with retarded and advanced interactions is investigated.
\end{abstract}


\section*{Introduction}
 One of the unsolved problems in theoretical physics is
 Hamiltonian and quantum description of discrete relativistic systems
 with retarded and advanced interactions.
 Such systems result from exclusion of field freedom degrees in
 field theories:
  while expressing fields by sources in classical field theories,
  or performing integration by fields in generating functionals
  of quantum field theories (carried out at definite boundary conditions)
  finite-dimensional theories arise, describing the motion of sources.

 Field exclusion in classical electrodynamics leads
 to Wheeler-Feynman system \cite{Wheeler}
 --- a relativistic theory with finite number of freedom degrees,
 whose action is Poincare-invariant functional
 of world lines of charges.
 This theory is equivalent to classical electrodynamics under
 the following boundary condition:
 field emitted by a charge is completely absorbed by other charges,
 so that outside some sphere (of radius greater than the Universe one)
 the resulting field of charges vanishes.
 The given formulation of electrodynamics has the following properties:

    1. In Wheeler-Feynman electrodynamics
      a field created by a point charge is a half-sum
      of retarded and advanced potentials
      (dislike the standard approach, which uses only the retarded potentials).
      Consequently, the time reversion symmetry in this theory
      evinces not only in the action,
      but also in solutions of equations of motion.

    2. T-nonsymmetric effects like radiative damping
      arise in multibody problems and have thermodynamic nature
      \cite{Wheeler}.

    3. As the classical theory has finite number of freedom degrees,
      one should expect that the corresponding quantum theory
      has no divergencies.

      Canonical quantization of Wheeler-Feynman electrodynamics
      has not been performed, because Hamiltonian formulation
      of this theory was unknown.
      Moreover, Lagrange equations of motion in Wheeler-Feynman
      electrodynamics are not ordinary differential equations,
      but belong to a poorly
      investigated class of functional equations
      (differential equations with deviating arguments \cite{difotkl}),
      that has no elaborate approaches to solution:
      neither general analytical methods, nor steady numeric techniques.
      Existence and uniqueness of solutions for this class of equations
      have not been studied thoroughly.
      That's why a general structure of solutions of classical electrodynamics
      in Wheeler-Feynman formulation is still unknown.

 Particular results obtained in this domain are:
   solution uniqueness theorem for large distances between charges
     \cite{Driver};
   exact solution of the problem for circular motion of charges
     in the theory with retarded and advanced interactions
     \cite{Schild},
     and the theory with retarded interaction and radiative 
reaction \cite{Rivera};
   exactly solvable modification of the problem \cite{Hill0},
     with one charge influenced by retarded interaction,
     and the other --- by advanced one;
   numerical solution of 1-dimensional 2-body problem \cite{Baeyer},
     extended to velocities $v<0.9545c$.

 Here we will continue the solution of the last problem
 to greater velocities.
 Besides that we will find new solutions in the range of $v<0.9545c$ 
 (along with those in \cite{Baeyer}).
 Hamiltonian formulation of the theory constructed in \cite{fw_old}
 is used for the problem solution.

 The paper has the following structure:
 the first section describes the methods under the use,
 the second one presents the obtained results.
 In three appendices boundary effects and limiting regimes are examined,
 and also a list of additional questions
 interesting for further investigation is given.


\section{Methods}\label{met}

\ortograph{0.}
 We will consider
 one-dimensional scattering of two relativistic particles
 of equal charges and masses.
 The methods used allow to study scattering of particles of different masses,
 however, the case of identical particles is of the most interest
 due to additional symmetries the problem acquires.

 We use the following system of units:
 light velocity $c=1$, classical radius of a particle $e^{2}/mc^{2}=1$
 (all distances are measured in classical radii).

\ortograph{1.}
 A method described in \cite{fw_old} is used for the problem solution.
 The clue idea of the approach is
 a choice of special parametrization of particles world lines,
 in which the problem gets the following formulation \bfref{f1}.
 Let's consider a system of particles $x_{n},y_{n}$,
 moving in 2-dimensional space-time so,
 that they are always located in vertices of a polygonal line
 assembled of light rays.
 It's enough to set a motion of particles $x$,
 and define $y$ trajectories as
 $y^{+}_{n}(\tau)=x^{+}_{n}(\tau)$,
 $y^{-}_{n}(\tau)=x^{-}_{n+1}(\tau)$
 (in light coordinates $x^{\pm}=x^{0}\pm x^{1}$).
 The motion is defined by a system of differential equations for
 $x^{\pm}_{n}(\tau)$ (simultaneous by $\tau$),
 which is equivalent to the original one with deviating arguments.
 Then additional conditions are imposed onto the particles trajectories,
 ensuring their sewing together to
 the whole paths of $x$ and $y$ particles.
 Therefore, this approach reduces the problem to
 a set of ordinary differential equations,
 that defines particles evolution,
 and a set of equations for initial conditions and evolution time of the form
 $F(X)=0$,
 that provides trajectories sewing together.

\figl{f1}{~\epsfysize=6cm\epsfxsize=5cm\epsffile{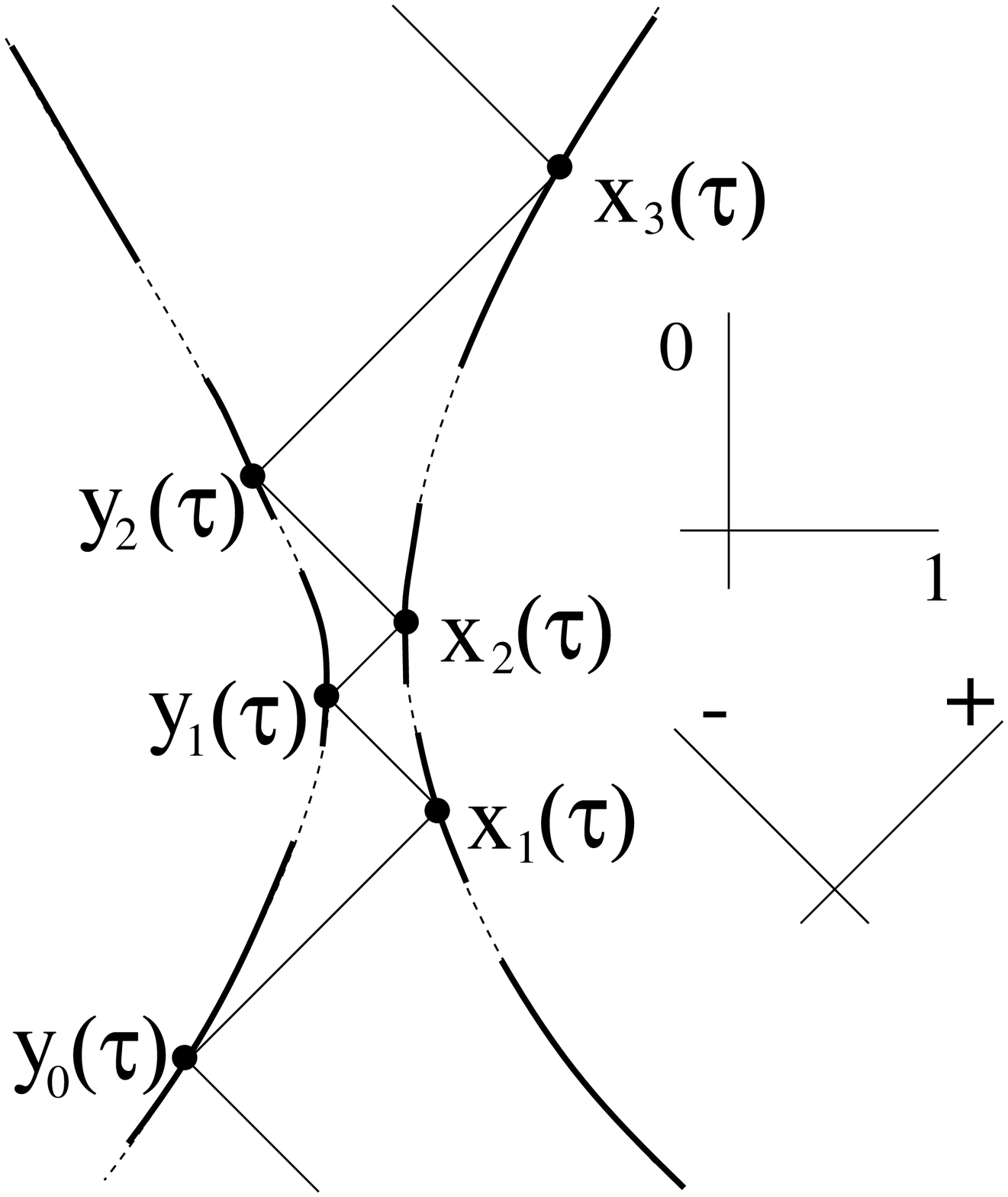}}
{Light stairway parametrization.}

 It's possible to restrict calculations to a finite number of trajectories
 $x_{n},\ n=1...N$.
 This possibility follows from exponential increase of
 distance between particles with large $n$:
 $|x_{n}-y_{n}|\sim q^{2n},\ q=(1+v)/(1-v),$
 $v$~---~the~velocity of the particles in center-of-mass frame (CMF);
 and starting from some $N$, their interaction becomes negligible.
 As $q\to\infty$ while $v\to1$,
 for large velocities few ($N=2,3$) steps of the light stairway are sufficient
 to move the particles out of the interaction region.
 Thus, this method is especially convenient in ultra-relativistic case
 (for small velocities more equations should be considered).

 The motion is defined by a system of Hamiltonian equations
 for coordinates and momenta:
\begin{eqnarray}
&&\dot x^{\pm}_{n}={{\d H}\over{\d p^{\pm}_{n}}},\quad
\dot p^{\pm}_{n}=-{{\d H}\over{\d x^{\pm}_{n}}}\label{difur}
\end{eqnarray}
 with Hamiltonian:
\begin{eqnarray}
&&H(x,p)=(1\;0)\;g_{N}...g_{1}\left(\begin{array}{c}1\\0\end{array}\right),\quad
g_{n}=\left(\begin{array}{cc}r_{n}^{+}r_{n}^{-}-1&r_{n}^{+}\\
-r_{n}^{-}&-1\end{array}\right),\nn\\
&&r_{n}^{\pm}=2\left({{p^{\pm}_{n}}\over{m}}
+{{1}\over{x^{\pm}_{n+1}-x^{\pm}_{n}}}\right),\
n=1...N-1,\quad r_{N}^{\pm}=2p^{\pm}_{N}/m.\label{rp}
\end{eqnarray}
 The Hamiltonian is a polynomial function of $r_{n}^{\pm}$
 (linear for each $r_{n}^{\pm}$).
 It's convenient to rewrite the equations completely in terms of $(x,r)$,
 see \cite{fw_old}.

 The Hamiltonian is a Dirac's constraint \cite{Dirac},
 that briefly speaking means the following.
 As the Hamiltonian is conserved in evolution:
$$\dot H=\dot x^{\pm}_{n}\;{{\d H}\over{\d x^{\pm}_{n}}}+
\dot p^{\pm}_{n}\;{{\d H}\over{\d p^{\pm}_{n}}}=0,$$
 phase trajectories lie on the surface $H=Const$.
 In Wheeler-Feynman electrodynamics only those trajectories,
 that lie on the surface $H=0$, correspond to a physically 
meaningful evolution.
 This constraint, causing the fact that not all momenta are independent,
 arises due to parametrical invariance of the action
and is typical for all relativistic theories.
 {\it E.g.} in a theory of free relativistic particle 
 analogous constraint has a form of a ``mass shell condition''
 $p^{2}-m^{2}=0$,
 identically satisfied on all trajectories, because
$p_{\mu}=m\dot x_{\mu}/\sqrt{\dot x^{2}}$ by definition.

 The sewing conditions have the form
\begin{eqnletter}
 \label{cont}
&&x_{n}^{\mu}(T)=x_{n+1}^{\mu}(0)\quad n=1...N-1\ ,\label{conta}\\
&&p_{n}^{\mu}(T)=p_{n+1}^{\mu}(0)\quad n=2...N-2,\quad \mu =+,-\ ;
\label{contb}\\
&&p_{1}^{+}(T)=p_{2}^{+}(0)-m/2\cdot(x_{2}^{+}(0)-x_{1}^{+}(0))^{-1},\nn\\
&&p_{N-1}^{-}(T)+m/2\cdot(x_{N}^{-}(T)-x_{N-1}^{-}(T))^{-1}=p_{N}^{-}(0),\nn
\end{eqnletter}
 for $n=2...N-2$ finite values of $(x_{n}^{\mu},p_{n}^{\mu})$
 coincide with initial values of these quantities for the next trajectory\relax
 \footnote{\relax
   For $n=1,N-1$ two $p_{n}$-conditions disappear, see \cite{fw_old}.
   Also, in the action transformation (2)$\to$(4) in \cite{fw_old},
   carried out for finite trajectories,
   off-integral terms appear, resulting to asymptotically vanishing corrections
   to the remaining two conditions.}.
 The quantities $(x_{n}^{\mu},p_{n}^{\mu})_{T}$,
 determined by solving differential equations (\ref{difur}),
 are functions of initial data $(x_{n}^{\mu},p_{n}^{\mu})_{0}$
 and ``time'' $T$.
 So (\ref{cont}) is a system of nonlinear equations on initial data and $T$.

 It has been shown in \cite{fw_old}
 that conditions (\ref{cont}) alone do not provide
 smoothness of trajectories sewing. 
 One more condition comes from the action minimum principle:
\begin{eqnarray}
&&u^{y}_{n}(T)=u^{y}_{n+1}(0),\ \mbox{где}\
u^{y}_{n}=\left({{dy_{n}^{-}}\over{dy_{n}^{+}}}\right)^{1/2}
={{\Psi_{n}^{1}}\over{\Psi_{n}^{2}}},\
\Psi_{n}=g_{n}...g_{1}\left(\begin{array}{c}1\\0\end{array}\right),\label{cont2}
\end{eqnarray}
 which is equivalent to smoothness of
 the $y_{n}$ and $y_{n+1}$ trajectories sewing together.
 This condition should be imposed for some $n=1...N-2$,
 and together with (\ref{cont}) it will provide smoothness of
 sewing together all $x_{n},y_{n}$ trajectories
 (except of the boundary ones: $x_{1,2},\ x_{N-1,N}$, 
  see Appendix~\ref{BoundaryEffects}). 

 Thus, $(4N-4)$ equations (\ref{cont}),(\ref{cont2}) and $H=0$ are imposed on
 $(4N+1)$ unknowns $(x_{n}^{\pm}(0),p_{n}^{\pm}(0),T)$.
 Among the remaining 5 freedom degrees,
 4 correspond to trivial transformations of solutions:
 \begin{itemize}
   \item 2 translations of solutions:\\
$x^{\pm}\to x^{\pm}+c^{\pm}$;
   \item Lorentz transformations:\\
$x^{+}\to x^{+}c,\ x^{-}\to x^{-}/c;\quad p^{+}\to p^{+}/c,\ p^{-}\to p^{-}c;$
   \item reparametrization (shift along the trajectories):\\
$(x,p)^{\pm}_{\tau}\to (x,p)^{\pm}_{\tau+c}$;
 \end{itemize}
 and one freedom degree corresponds to trajectories deformation
 in variation of relative velocity of the particles.

\ortograph{2.}
 The trivial freedom degrees can be eliminated,
 {\it e.g.} setting $x_{1}^{\pm}(0)=0,\ r^{-}_{1}(0)=Const$
 --- a value of this constant can be taken from Coulomb approximation
 (see below) and fixed in all further considerations.
 Applying translations and Lorentz transformation to a solution found,
 one can move it to another reference frame,
 {\it e.g.} CMF.

 In order to fix reparametrization freedom degree,
 let's add one more equation:
 $(x_{1}^{+}-x_{1}^{-})(0)=(x_{N}^{+}-x_{N}^{-})(T)$ in CMF,
 geometrical interpretation of which is clear from \fref{f3}.
 Amongst all of parametrization fixing constraints
 this one is especially convenient, because
 it prevents a solution from ``slipping off'' along a trajectory
 and ensures that boundary points are distant from interaction region,
 both in the past, and in the future.

 \figl{f3}{~\epsfysize=6cm\epsfxsize=3cm\epsffile{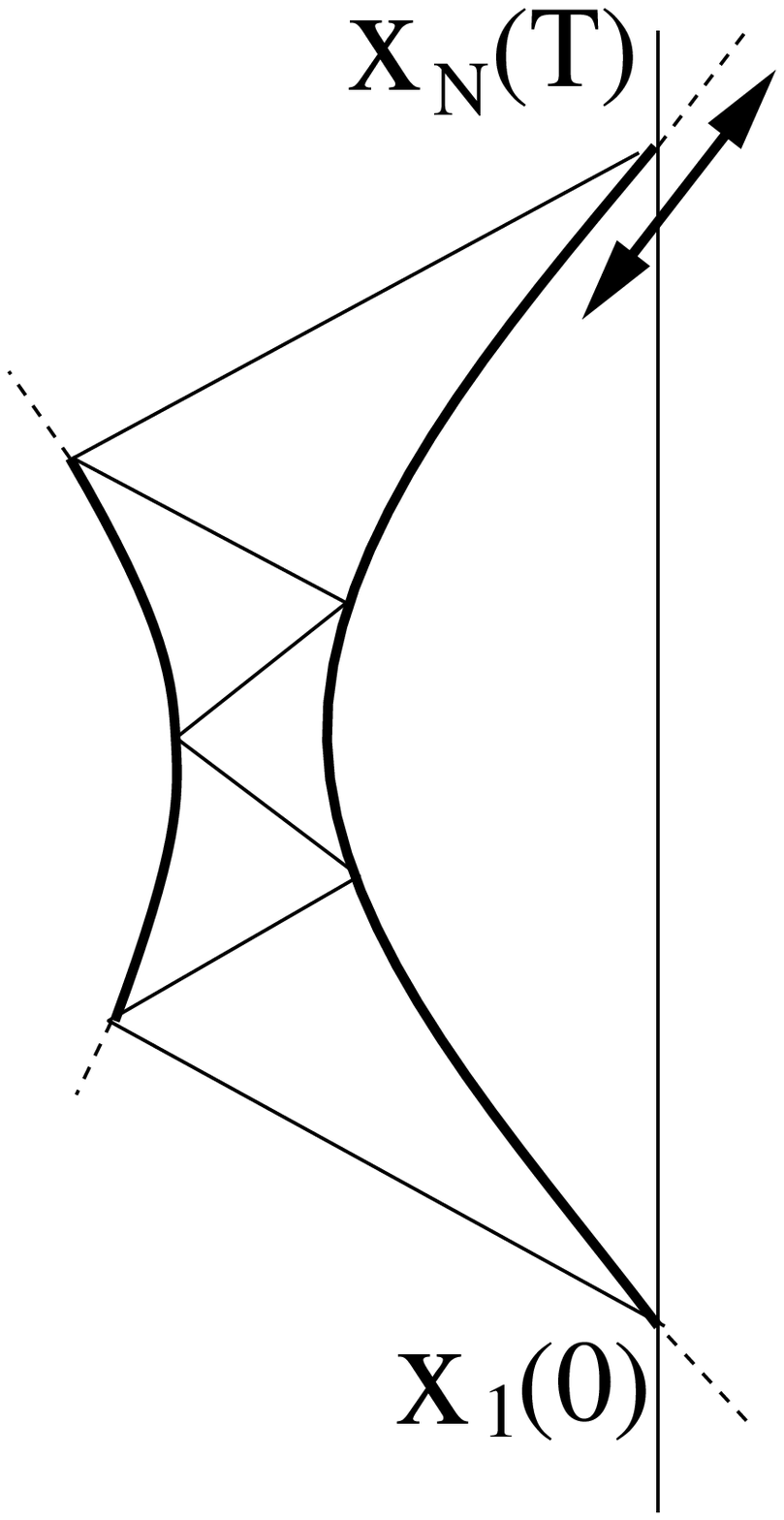}}
 {Fixing reparametrization freedom degree.}

One more variable should be fixed in the problem,
 {\it e.g.} $r^{+}_{1}(0)$, it will control particles velocities.
 The expression for $v$
 (initial velocity of particles $x_{1},y_{1}$ in CMF)
 has the form: $v=2(r_{1}^{+}(0)r_{1}^{-}(0))^{-1}-1$, see \cite{fw_old}.
 It's convenient to introduce a parameter $\mu=(1-v)^{-1}$
 with range $[1,+\infty)$,
 and express $r^{+}_{1}(0)$ by it.

 Equation $H=0$ is linear for each $r_{n}^{\pm}$
 and can be solved explicitly ({\it e.g.} for $r_{2}^{-}(0)$).
 Thus, with the changes made,
 we have the system of $(4N-4)$ equations $F(\mu,X)=0$
 for $(4N-4)$ unknowns $X$,
 depending on one parameter $\mu$.

 Newton's method was applied to resolve this system.
 Coulomb (non-relativistic) solution of the problem at $v=0.5$ was used 
as a starting point, it has the form (in the designations introduced):

 {\footnotesize
  \begin{eqnarray}
    &&x^{1}(\tau)={{1}\over{2v^{2}}}\;\ch^{2}2v\tau,\quad
    x^{0}(\tau)={{1}\over{2v^{3}}}\;(2v\tau+\ch 2v\tau\sh 2v\tau),\nn\\
    &&x^{\mu}_{n}=x^{\mu}(\tau_{n}),\quad \tau_{n}=n-1-N/2\quad n=1...N,\nn\\
    &&u^{x}_{n}=\left({{\dot x_{n}^{-}}\over{\dot x_{n}^{+}}}\right)^{1/2}
    \quad n=1...N,\quad
    u^{y}_{n}=\left({{\dot x_{n+1}^{-}}\over{\dot x_{n}^{+}}}\right)^{1/2}
    \quad n=1...N-1,\nn\\
    &&r_{n}^{+}= -u^{x}_{n}-u^{y}_{n}\quad n=1...N-1,\quad
    r_{n}^{-}=-{{1}\over{u^{x}_{n}}}-{{1}\over{u^{y}_{n-1}}}\quad n=2...N,
    \label{ru}\\
    &&r_{N}^{+}= -u^{x}_{N},\quad r_{1}^{-}= -{{1}\over{u^{x}_{1}}}.\nn
  \end{eqnarray}
 }

 Then after a few iterations precise solution was found\relax
 \footnote{The condition $|F|<10^{-5}$ has been used as a finish criterion;
   while far from solutions $|F|\sim 10^{2}...10^{3}$.}.
 Then the $\mu$ parameter was increased by $\Delta\mu$,
 and the solution found was used as a starting point
 for new value of $\mu$.
 The step $\Delta\mu$ was chosen adaptively: it was
 \begin{itemize}
   \item decreased by 2 times, if convergency has been lost;
   \item increased by 2 times, if solution has been found;
 \end{itemize}
 so $\Delta\mu$ has been automatically kept in an optimal region\relax
 \footnote{Other optimization methods are described in \cite{fw_prog}.}.

\ortograph{3.}
 With the help of these methods the solution was continued up to
 velocity $v=0.937$.
 At this value of velocity Jacoby matrix of considering system
 degenerates (see \fref{g1}),
 and Newton's method becomes inapplicable.
 In fact, the solution extrapolation enables to ``jump over'' this point
 and continue the solution ($0$) behind it.
 However, vanishing of the Jacobian indicates a change of topology for the
 set of solutions,
 therefore the vicinity of this point should be studied by other methods.

 \figl{g1}{~\epsfysize=6cm\epsfxsize=6cm\epsffile{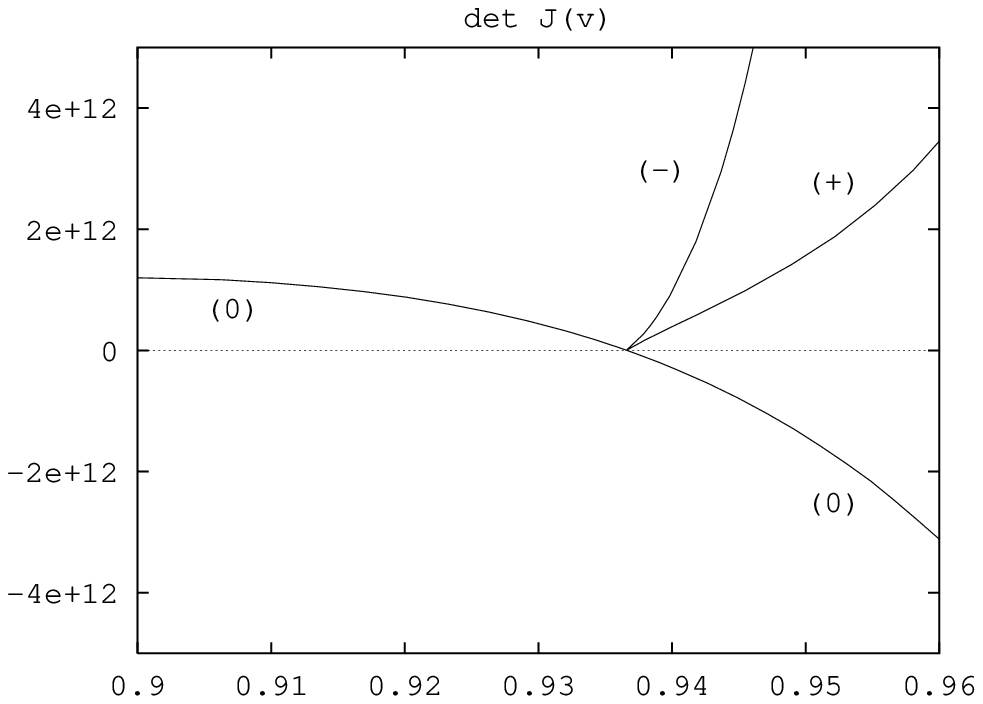}}
 {Jacobian dependence on velocity.}

\begin{figure}\label{g2}
   ~\epsfysize=6cm\epsfxsize=6cm\epsffile{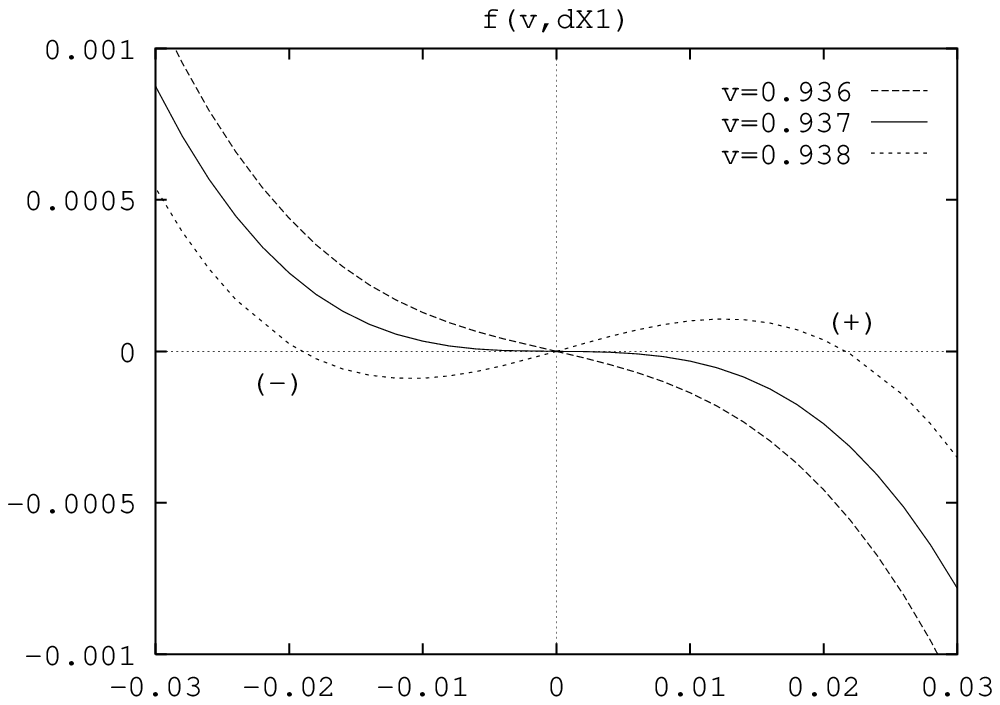}\hfill
   ~\epsfysize=6cm\epsfxsize=6cm\epsffile{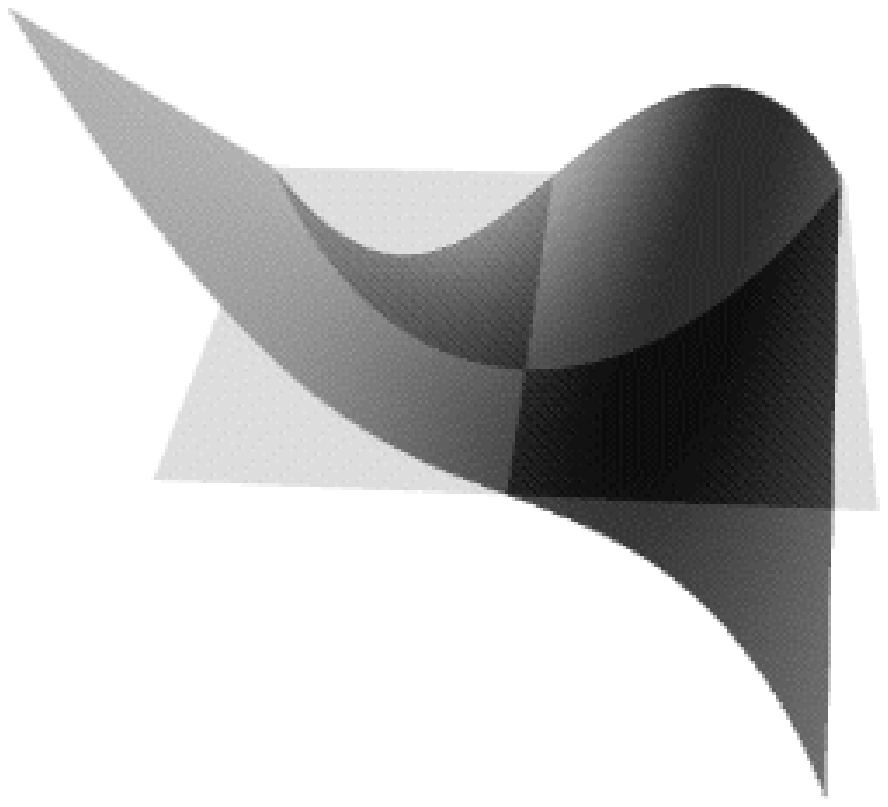}

   \vspace{0.3cm}
   \fignum Dependence $f(v,\Delta X_{1})$.
   In the vicinity of the critical point this function defines a surface,
   known in catastrophe theory as {\it Cayley surface} \cite{Francis}.
 \end{figure}

 Let's extract in the Jacoby matrix $J_{ij}=\d F_{i}/\d X_{j}$
 of size $K\times K,\ K=4N-4$,
 a submatrix $\tilde J$ of size $(K-1)\times(K-1)$,
 with determinant not-vanishing in the critical point.
 Place this non-degenerate block to bottom right matrix corner
 by renumerating the variables and equations.
 Now, fixing $(\mu,X_{1})$,
 it's possible to resolve $(K-1)$ equations
 $F_{i}(\mu,X_{1},X_{j})=0,\ i,j=2...K,$
 for $(K-1)$ unknowns $X_{j}$,
 using Newton's method.
 Non-degeneracy of $\tilde J$ guarantees that the solutions found are isolated:
 for fixed $(\mu,X_{1})$ in the vicinity of the solution $X_{j}$
 there are no other ones.
 Further, we get $X_{j}(\mu,X_{1})$ dependencies,
 carrying out step-by-step change of parameters $(\mu,X_{1})$,
 similar to the described above.
 Substituting these dependencies to the remaining equation,
 we get a function of two variables
 $f(\mu,X_{1})=F_{1}(\mu,X_{1},X_{j}(\mu,X_{1}))$,
 a behavior of which is shown on the \fref{g2}
 (in terms of variables $v,\ \Delta X_{1}=X_{1}-X_{1}^{0}$,
  where $X_{1}^{0}$ --- solution ($0$)).

 At $v<0.937$ the equation $f(\Delta X_{1})=0$ has a single solution.
 In the critical point a tangent to the graph at zero is horizontal
 (that's equivalent to $\det J=0$).
 At $v>0.937$ the tangent slope reverses its sign
 and two additional solutions appear.
 Thus, in the critical point bifurcation of solutions takes place:
 one solution splits into three.

 \vspace{2mm}\noindent

\vspace{0.2cm}\noindent
\begin{figure}\label{g3}
\hfill
\mar\epsfysize=3.2cm\epsfxsize=3.8cm\epsffile{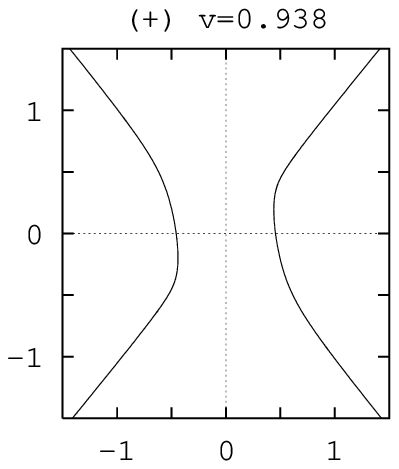}
\mar\epsfysize=3.2cm\epsfxsize=3.8cm\epsffile{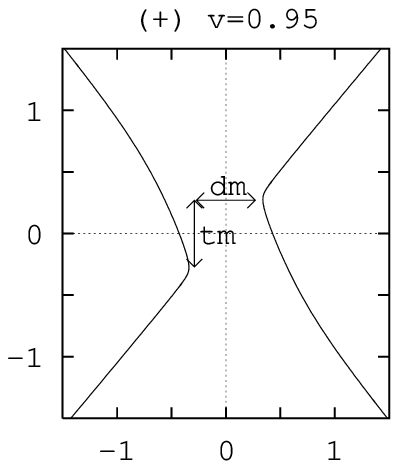}
\mar\epsfysize=3.2cm\epsfxsize=3.8cm\epsffile{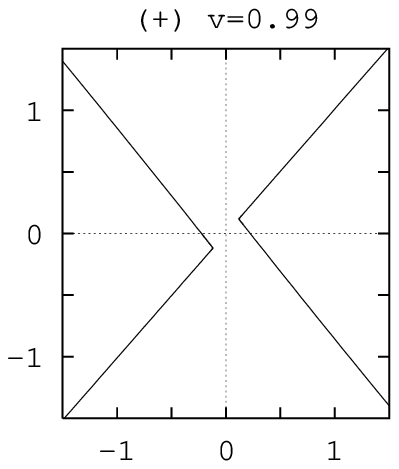}

\vspace{1mm}

\hfill
\mar\epsfysize=3.2cm\epsfxsize=3.8cm\epsffile{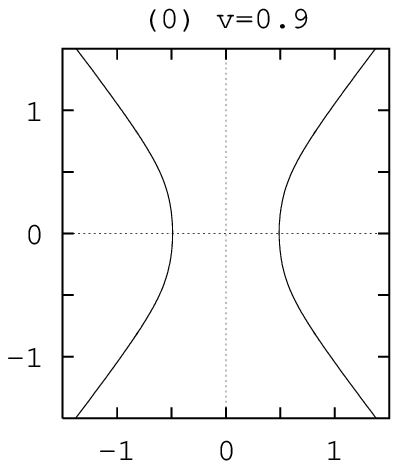}
\mar\epsfysize=3.2cm\epsfxsize=3.8cm\epsffile{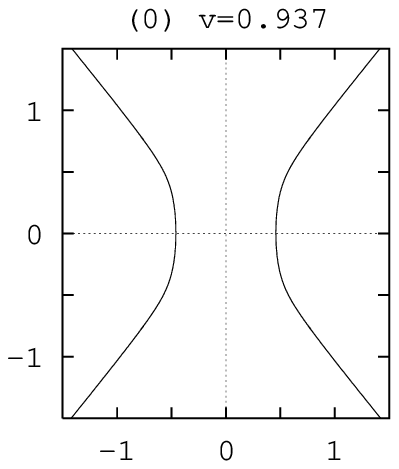}
\mar\epsfysize=3.2cm\epsfxsize=3.8cm\epsffile{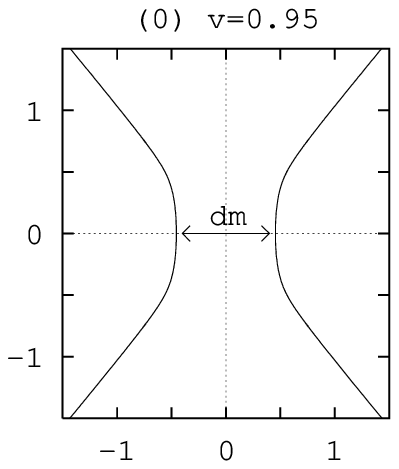}
\mar\epsfysize=3.2cm\epsfxsize=3.8cm\epsffile{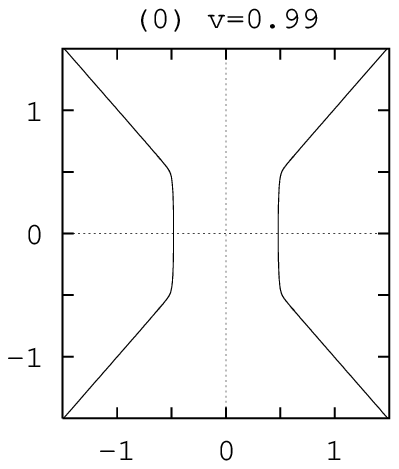}

\vspace{1mm}

\hfill
\mar\epsfysize=3.2cm\epsfxsize=3.8cm\epsffile{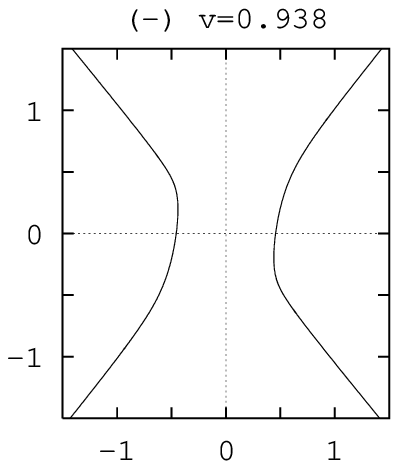}
\mar\epsfysize=3.2cm\epsfxsize=3.8cm\epsffile{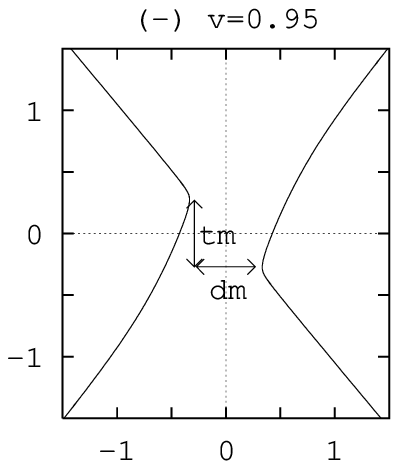}
\mar\epsfysize=3.2cm\epsfxsize=3.8cm\epsffile{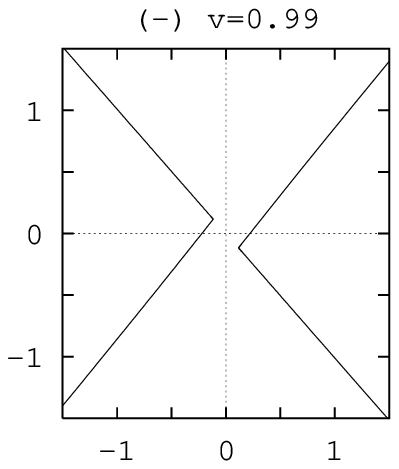}

\vspace{2mm}
\begin{center}
\fignum Trajectories shape.
\end{center}
\end{figure}

 \vspace{2mm}

 Positions of the additional solutions are determined using dichotomy.
 Taking $v$ far enough from the critical point
 one can continue the problem solution using Newton's method.


\section{The results}

\ortograph{1.}
 Shapes of trajectories corresponding to found solutions
 are shown in \fref{g3}.
 While passing critical point no irregularities of trajectories and graphs of
 velocities and accelerations are observed.

 For the main branch ($0$)
 trajectories are symmetric with respect to spatial and temporal reflections
 $P:\ x^{1}\to -x^{1}$ and $T:\ x^{0}\to -x^{0}$.

 For the additional branches ($\pm$) trajectories are not
 $P$- and $T$-symmetric,
 but transform to each other in these reflections.
 At the same time, these trajectories are $PT$-symmetric.

 This effect:
 violation of solutions symmetry for symmetrically stated problem
 is related with non-uniqueness of solutions.
 Indeed, a symmetry of equations of motion (and initial data)
 under reflections means only 
 that the reflected solution is also a solution (maybe another one),
 in other words, a set of solutions under reflections transits to itself.
 Only if a solution at the given initial data is unique
 ({\it e.g.} as for ordinary differential equations),
 one can conclude that the solution coincides with the reflected one,
 {\it i.e.} is symmetric.
 Thus, violation of $P,T$-symmetries for solutions is specific to
 differential equations with deviating arguments,
 which can have multiple solutions for the same initial data.

\begin{figure}\label{g4}
\mar\epsfysize=7cm\epsfxsize=7cm\epsffile{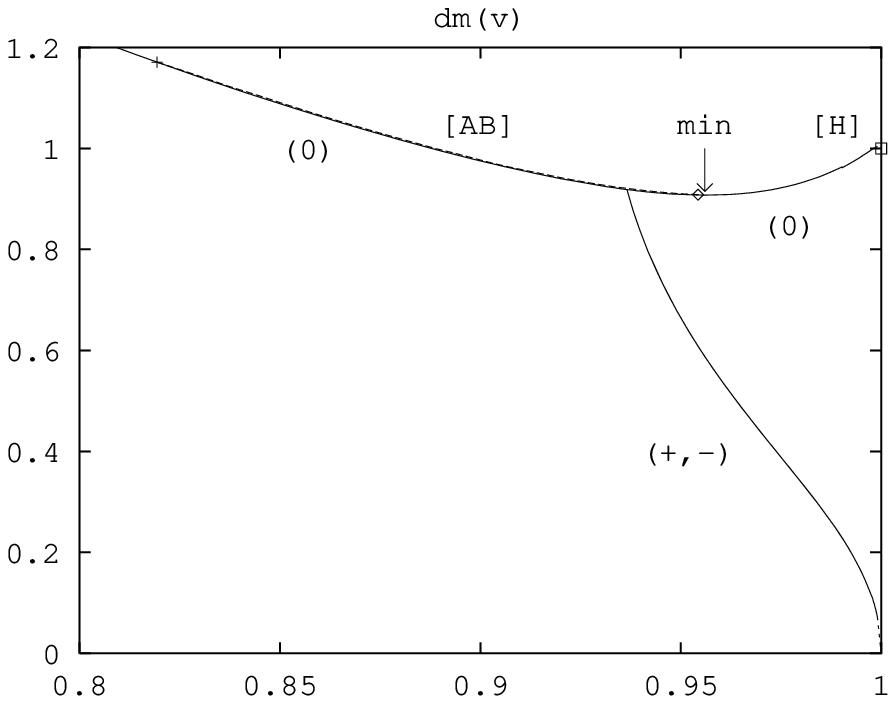}\hfill
\mar\epsfysize=7cm\epsfxsize=7cm\epsffile{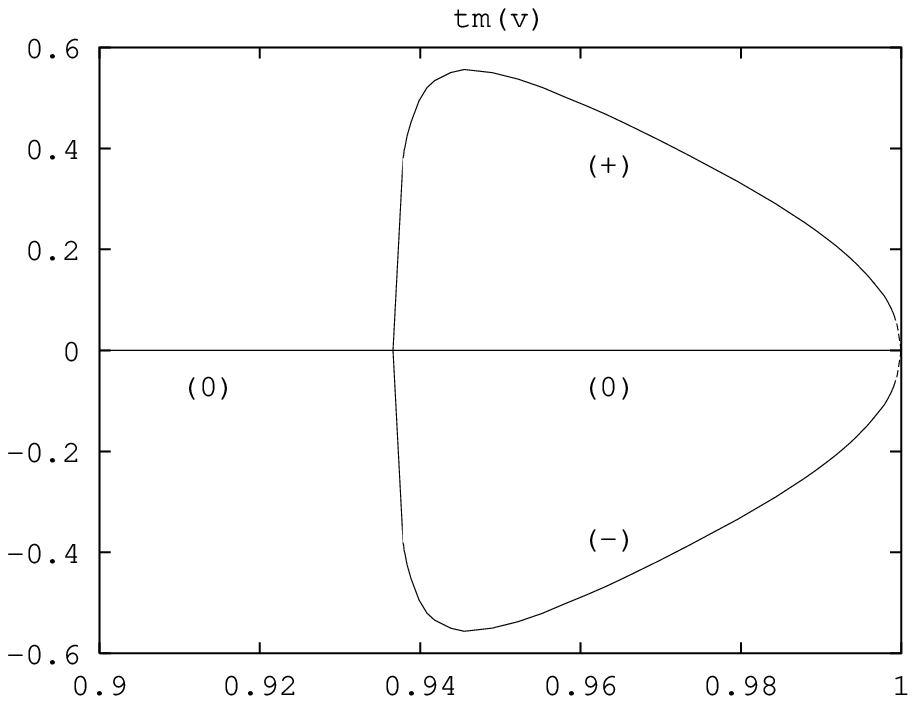}

   \vspace{0.3cm}
   \fignum $d_{m}(v)$ and $t_{m}(v)$ dependencies.
   [AB] --- Andersen -- von Baeyer solution,
   [H] --- Hill's solution.
 \end{figure}

\ortograph{2.}
 The dependencies of parameters $d_{m},t_{m}$ on $v$ are shown in \fref{g4}
 (the definitions of parameters see in \fref{g3}:
  $d_{m}=\min x^{1} -\max y^{1}$ in CMF,
  $t_{m}$ --- temporal distance between the points,
  where extrema are reached).

 The dependence $d_{m}(v)$ for the main branch of solution has minimum at
 velocity $v_{m}=0.956$, corresponding to $d_{m}=0.9075$.
 This dependence (up to velocity value close to $v_{m}$)
 has been found in Andersen and von Baeyer work \cite{Baeyer}.
 Overlaying the graph \cite{Baeyer} to the one obtained here,
we see their exact coincidence.

 \begin{quot}
   \noindent{\it Note.}
   The following algorithm has been used in \cite{Baeyer}
   for the problem solution.
   For a given path of particle $x$ the forces
   acting on it from a particle $y$ were calculated,
   assuming path of the $y$ is a {\it mirror reflected image} of path of 
the $x$.
   Then the accelerations were integrated,
   corrections to $x$ trajectories were found,
   and the described process was iterated.
   First of all,
   let's note that this algorithm considerably uses
   assumption about $P$-symmetry of the trajectories,
   therefore it is only capable to obtain symmetrical solutions
   corresponding to the main branch $(0)$.
   Then in order to find a solution the described process
   was performed at a fixed $d_{m}$ value,
   which was diminished while moving to the range of greater velocities.
   The iterations converged till $d_{m}=0.9077$
   (that corresponds to $v=0.9545$)
   and diverged for less $d_{m}$ values.
   This divergency was caused by proximity of
   a minimum of $d_{m}(v)$ dependence
   --- there are no solutions for $d_{m}<0.9075$.
 \end{quot}

\begin{figure}\label{g3a}
   \begin{center}

\mar~~~\epsfysize=4cm\epsfxsize=4cm\epsffile{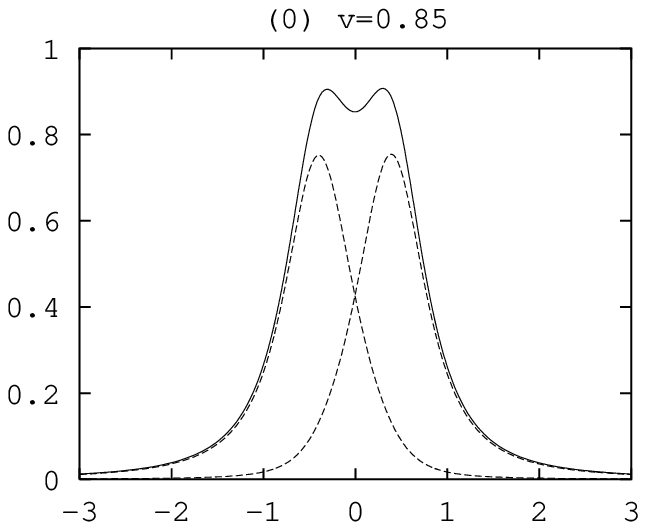}
\mar~~~\epsfysize=4cm\epsfxsize=4cm\epsffile{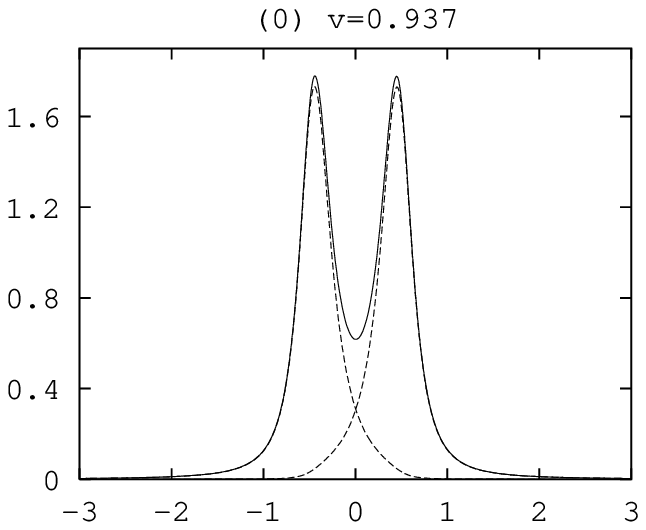}
\mar~~~\epsfysize=4cm\epsfxsize=4cm\epsffile{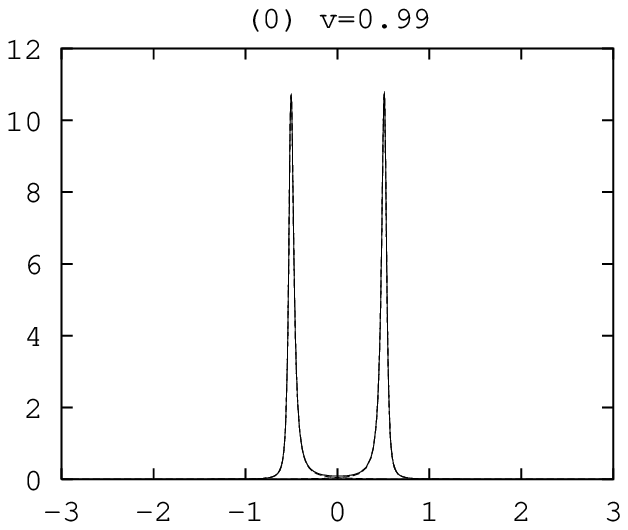}

     \vspace{2mm}
\mar~~~\epsfysize=4cm\epsfxsize=4cm\epsffile{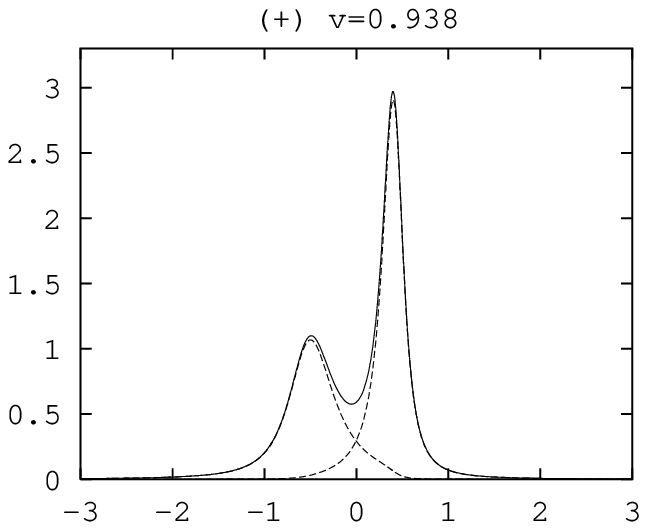}
\mar~~~\epsfysize=4cm\epsfxsize=4cm\epsffile{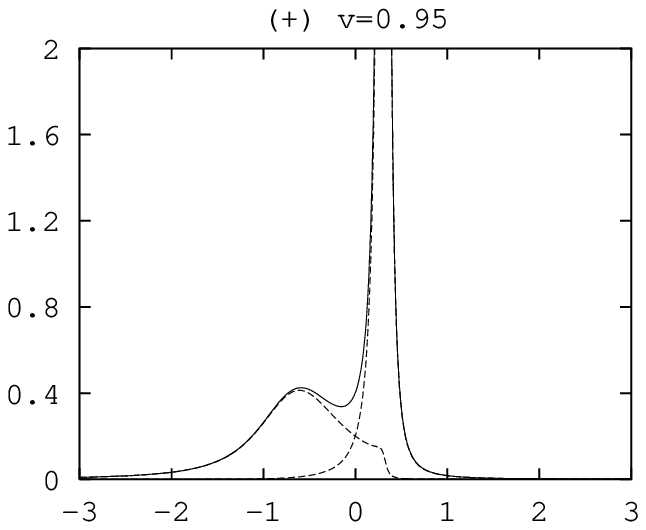}
\mar~~~\epsfysize=4cm\epsfxsize=4cm\epsffile{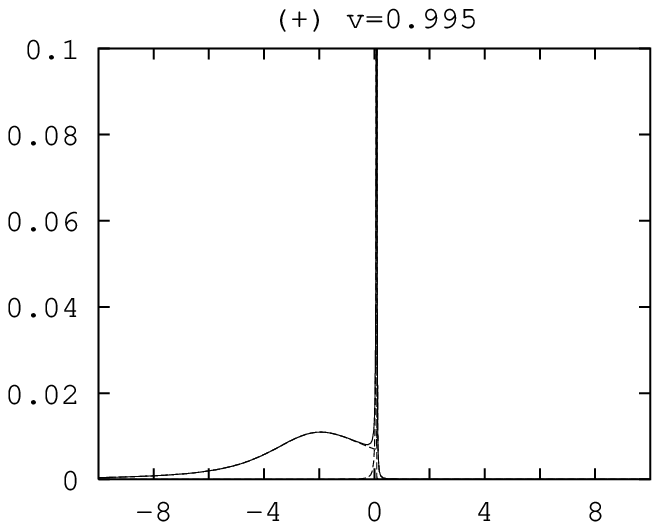}

   \end{center}

   \fignum Graphs of accelerations $d^{2}x^{1}/(dx^{0})^{2}$ in CMF.
   Dashed lines present contributions of retarded and advanced interactions.
 
 \end{figure}

 For $v\to1$ solution in the main branch tends to Hill's solution \cite{Hill},
 for which the minimal distance between particles in CMF is $d_{m}=1$.
 The trajectories are polygonal lines with vertices
 in points $(\pm0.5,\pm0.5)$, see \fref{g3}.
 There are $\delta$-like peaks on the accelerations graphs
 corresponding to these vertices (see top right graph in \fref{g3a}),
 one of them corresponds to retarded,
 and another one --- to advanced interactions.

 For $(\pm)$ solutions $d_{m},t_{m}\to0$ for $v\to1$.
 The trajectories in CMF tend to light rays emitted from the frame origin.

 \noindent

 \noindent

 \begin{figure}\label{g5}

   \hfill
\mar\epsfysize=4.3cm\epsfxsize=4.3cm\epsffile{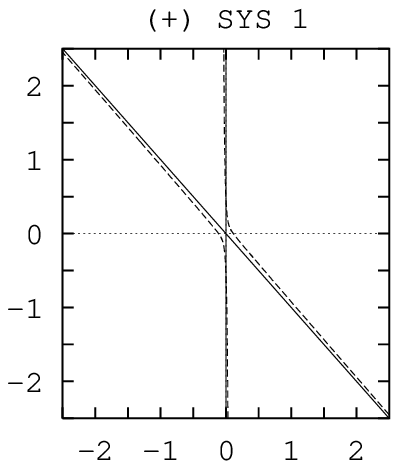}
\mar\epsfysize=4.3cm\epsfxsize=4.3cm\epsffile{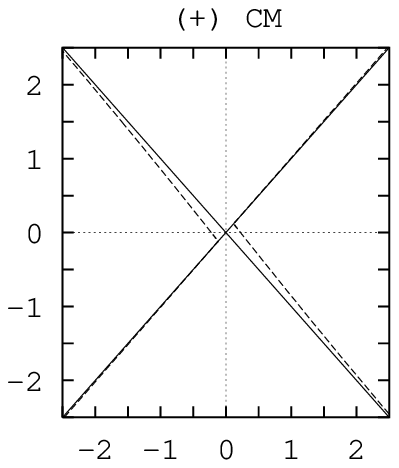}
\mar\epsfysize=4.3cm\epsfxsize=4.3cm\epsffile{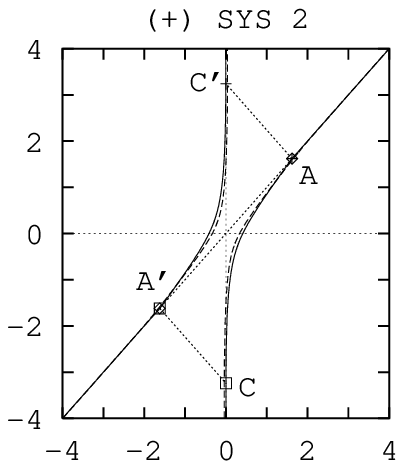}

\hfill ~\epsfysize=4.3cm\epsfxsize=4.3cm\epsffile{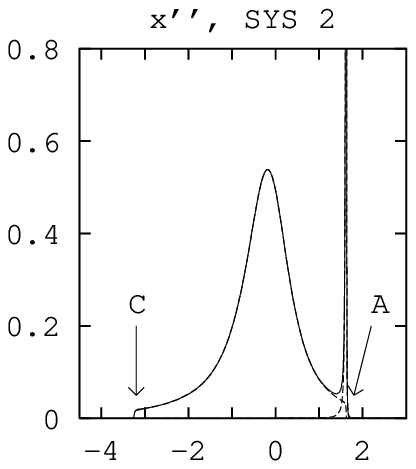}

  \vspace{-3cm}
   \parbox[c]{8cm}
   {\fignum
    On the top: $(+)$ solutions in various reference systems
    (dashed lines: $v=0.95$, solid lines: $v\to1$.)
    On the bottom: acceleration of particle $x$ in the system 2.
   }
   \hfill

 \end{figure}

 \vspace{2.5cm}

 It's interesting to investigate limiting shapes of the trajectories
 in reference systems,
 where one of the particles has zero velocity at $t\to\pm\infty$.
 There are two such systems (see \fref{g5}),
 in one of them solution collapses to the frame origin, as in CMF.
 In the other system another regime is established.
 The trajectories have small fractures (breaks of smoothness),
located in the points $A,A'$.
 These fractures are related with $\delta$-like acceleration peaks,
 which correspond to the interactions of one type
 (in \fref{g5} -- retarded for $x$, and advanced for $y$).
 Interaction of another type has smooth dependence on time
 in ranges $CA,A'C'$ and defines trajectories shape in these intervals.
 It's possible to give analytical description of this shape,
 see Appendix \ref{UltimateModes}.

\ortograph{3.}
 Poincare-invariance of action \cite{fw_old} follows to appearance
 of Noether's motion integrals, {\it i.e.} to the conservation of
 \begin{eqnarray}
   &&\mbox{translations generators}\ P^{\pm}=\sum_{i}p_{i}^{\pm}\label{id}\\
   &&\mbox{and boosts generator}\ M=\sum_{i}x_{i}^{+}p_{i}^{+}-x_{i}^{-}p_{i}^{-}.
 \nn\end{eqnarray}
 In this approach momenta $p_{i}^{\pm}$ are expressed
 via the $r_{i}^{\pm}$ values,
 which are uniquely defined by velocities $\dot x_{i}^{\pm}$,
 and the $(x_{i+1}^{\pm}-x_{i}^{\pm})^{-1}$ values,
 defined by distances between interacting particles, see (\ref{rp}),(\ref{ru}).
 The terms $(x_{i+1}^{\pm}-x_{i}^{\pm})^{-1}$ 
 correspond to interaction contribution to the integrals of motion,
 which is present in a standard approach \cite{Wheeler}
 (emitted but still not absorbed field).
 The difference from the standard approach is that
 integrals of motion (\ref{id}) comprise
 a sum over a sequence of points on the trajectories,
 linked by light stairway,
 which propagates both to the future and to the past of the system.

 Conservation of integrals (\ref{id})
 has been used for the control of solutions accuracy in numerical analysis,
 see \cite{fw_prog} for details.

\ortograph{4.}
 The solutions $X(\mu)$ found here were put in the Internet
 for the convenience of readers and further problem study.
 \begin{center}
   {\tt http://viswiz.gmd.de/\~{}nikitin/fw\_{}data/node1.html} .
 \end{center}


\section*{Conclusion}
 The analysis of 1-dimensional scattering of 2 charges of equal masses
 in Wheeler-Feynman electrodynamics has revealed splitting of solutions:
 the scattering is uniquely defined by asymptotic velocities of the
 charges for $v<0.937c$;
 for $v>0.937c$ there are 3 different solutions.
 Mirror reflection symmetry is violated in the splitting:
 one of three solutions is $P$-symmetric,
 the other two are not,
 but transit to each other under $P$-reflection.
 For limits of solutions at $v\to c$ analytical expressions have been derived.

\paragraph*{Acknowledgments.}
 The authors thank S.N.Sokolov and G.P.Pronko for useful discussions.
 The work has been carried out at
 Institute for High Energy Physics (Protvino, Russia) and
 National Research Center for Information Technology (Bonn, Germany)
 and has been partially supported by
 RFBR 96-01-01273 and INTAS 96-0778 grants.

\baselineskip=0.4\normalbaselineskip\footnotesize


\appendix
\section{Boundary effects}\label{BoundaryEffects}
\ortograph{1.}
 Sewing (\ref{cont}),(\ref{cont2}) leads to appearance of
fractures on the boundary parts of trajectories.
 A physical reason of this effect is the following.

 The equations of motion \cite{fw_old} have been derived
 minimizing the action, defined for finite trajectories, see \fref{f2}.
 Due to this fact the interaction between particles $y$ and $x$
 switches on/off instantaneously,
 potential of $y$ particle field has breaks at the points $x_{2},x_{N}$.
 These breaks correspond to $\delta$-like Lorentz force $F$,
 which causes trajectories fractures at the points $x_{2},x_{N}$.
Breaks of velocity at $x_{2},x_{N}$ leads to breaks
of second derivatives at $y_{2},y_{N-1}$, and so on.
 Coefficient at $\delta$-function in $F$,
 which determines amplitude of these breaks,
 is proportional to $\eta^{-1}$ and vanishes while $\eta\to\infty$
 ($N\to\infty$ or $v\to1$).
 In the examined range of velocities the
 effects caused by instantaneous switching of interaction
 appear to be small
 (a break of velocity in fracture points is
  $\Delta v<5\cdot10^{-4}$ for $v>0.9$, $\Delta v<5\cdot10^{-5}$ for $v>0.95$,
  see \cite{fw_prog}).

 \figl{f2}{~\epsfysize=6cm\epsfxsize=4cm\epsffile{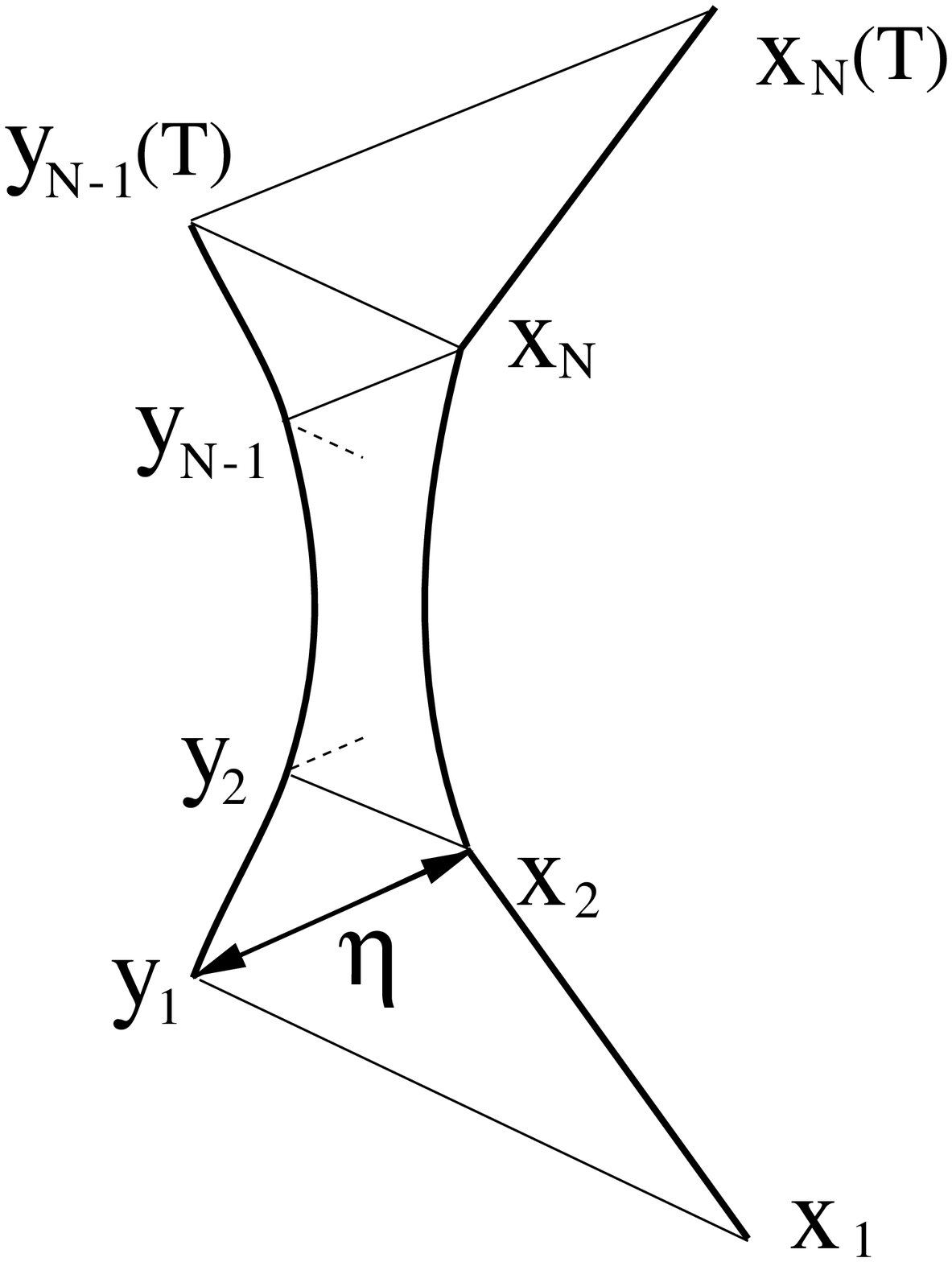}}
 {}

\ortograph{2.}
 Other schemes of interaction switching 
 ({\it e.g.} continuation of trajectories by infinite straight lines)
 are able to remove the fractures.
 However, these techniques are less logical,
 as they do not follow from the action minimum principle
 for finite trajectories.
 Consideration of infinite trajectories in stairway parametrization
 is related with additional difficulties, see \cite{fw_old}.

\vspace{5mm}
\ortograph{3.}
 For $N=3$ only the last 2 equations remain in (\ref{contb});
 in (\ref{cont2}) $n=1$.
 The trajectories $y_{1},y_{2}$ are sewed together smoothly,
 fractures appear in sewing $x_{1},x_{2},x_{3}$ trajectories together.

 For $N=2$ conditions (\ref{contb}) are absent,
 instead of (\ref{cont2}) one should use a condition
 \begin{eqnarray}
   &&u^{x}_{1}(T)-u^{x}_{2}(0)={{u^{x}_{1}(T)\;u^{x}_{2}(0)}\over
   {(x_{2}^{-}-x_{1}^{-})(T)}}+{{1}\over{(x_{2}^{+}-x_{1}^{+})(0)}},\quad
   u^{x}_{1}=-{{1}\over{r_{1}^{-}}},\ u^{x}_{2}=-r_{2}^{+}
   \qquad(\ref{cont2}')\nn
 \end{eqnarray}
 --- $x_{1}$ and $x_{2}$ trajectories are sewed together with a fracture,
 caused by action of $\delta$-like Lorentz forces
 from bounds of $y_{1}$ trajectory.

\section{Limiting regimes}\label{UltimateModes}
\begin{figure}\label{proc}
a)~~\epsfysize=3.5cm\epsfxsize=3.5cm\epsffile{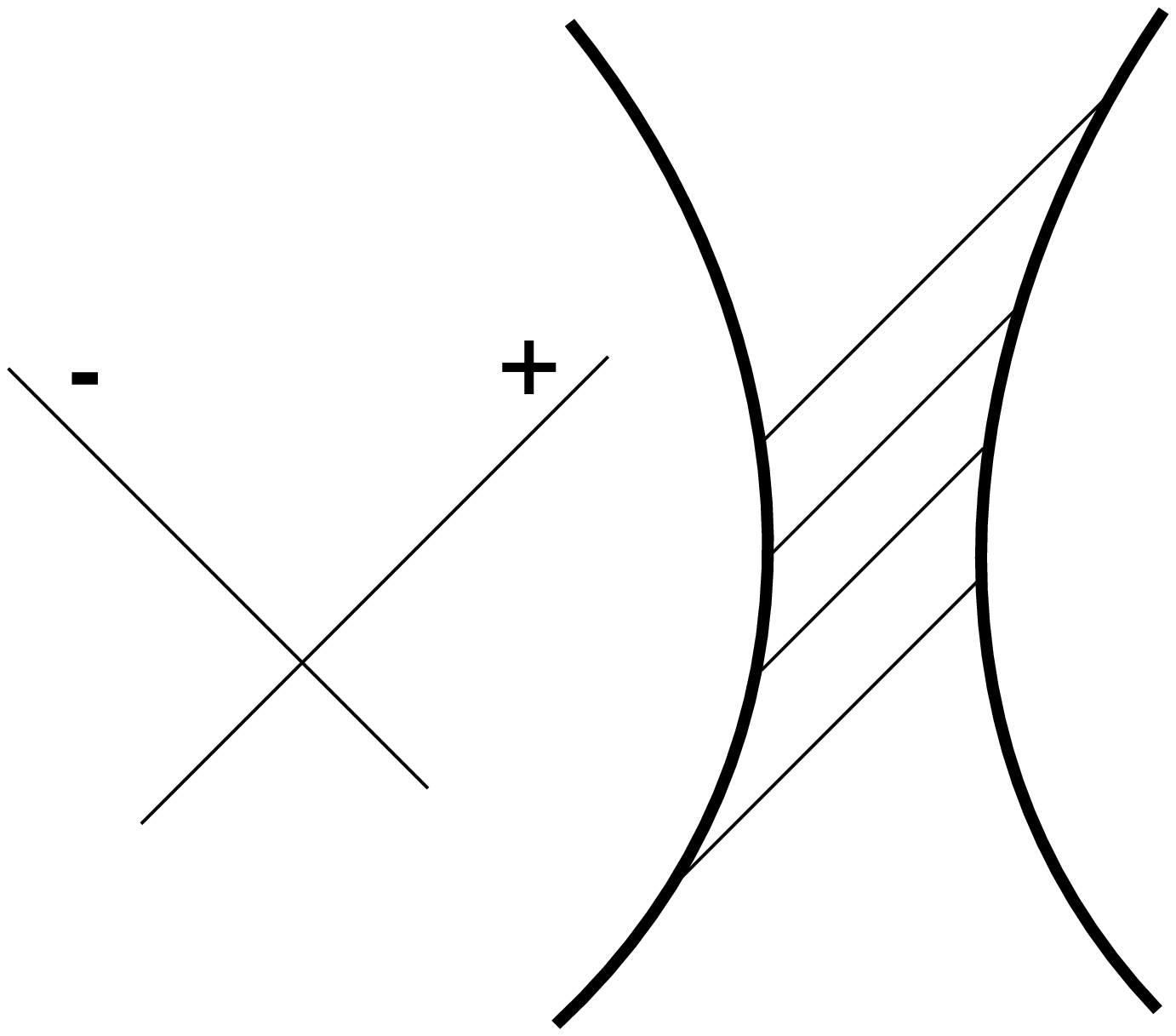}
~~~~b)~~\epsfysize=3.5cm\epsfxsize=3.5cm\epsffile{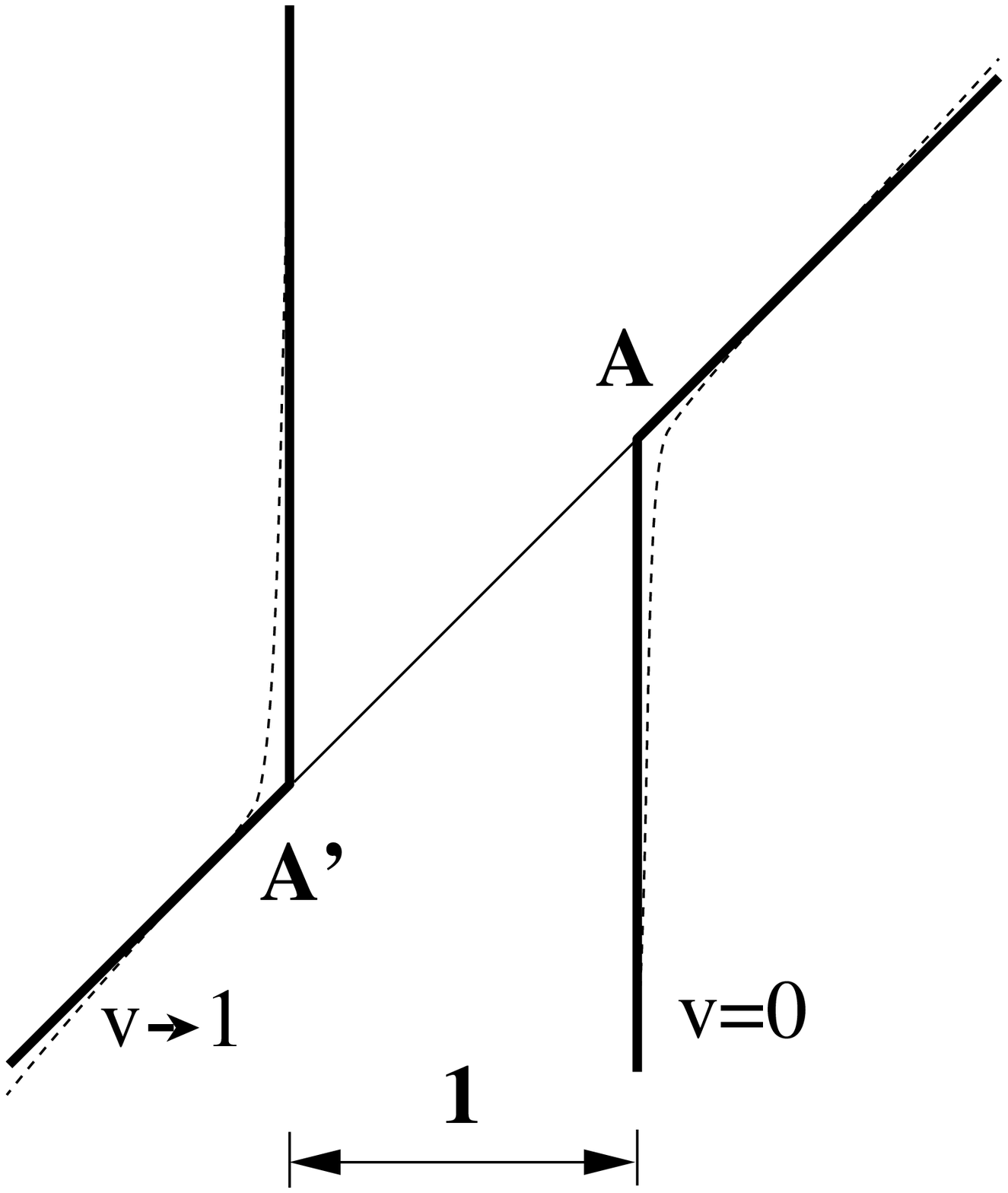}
~~~~c)~~\epsfysize=3.5cm\epsfxsize=3.5cm\epsffile{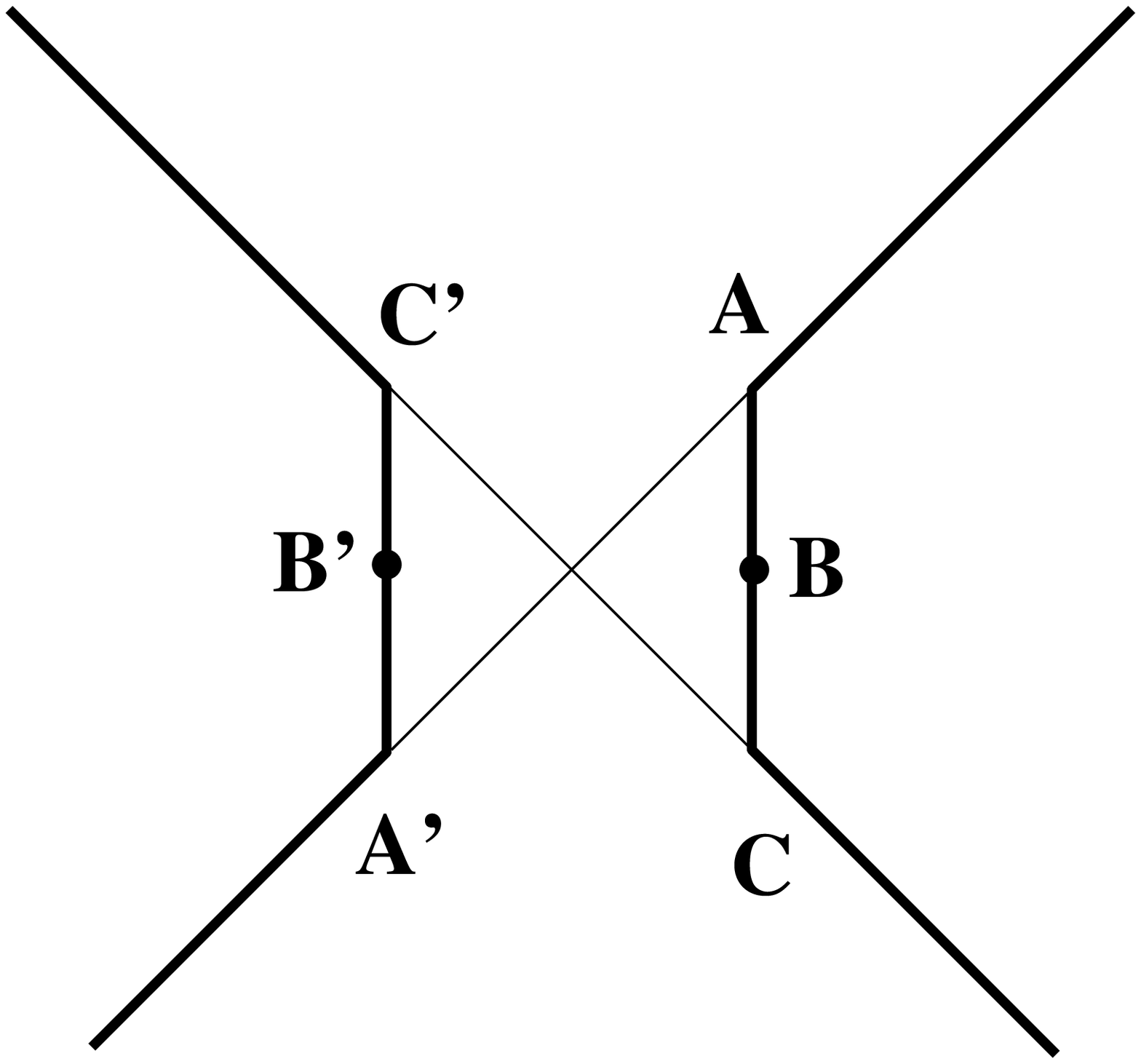}

   \vspace{0.3cm}
   \fignum 
 \end{figure}

\ortograph{1.}
 A problem of particles motion, influenced by interaction of one type
 ({\it e.g.} retarded for $x$ and advanced for $y$)
 is exactly solvable \cite{Hill0}.
 Indeed, in light coordinates
 such a motion is described by differential equations,
 simultaneous by $\tau=x^{-}$, see \fref{proc}a.

\vspace{3mm} 

 \vspace{1cm} 
 The equations have the form:
 \begin{eqnarray}
   &&\ddot x^{+}={{4{\dot y}^{+}{\dot x}^{+}{}^{3/2}}\over{(x^{+}-y^{+})^{2}}},
   \quad 
   \ddot y^{+}=-{{4{\dot x}^{+}{\dot y}^{+}{}^{3/2}}\over{(x^{+}-y^{+})^{2}}},
   \quad x^{-}=y^{-}=\tau,\nn
 \end{eqnarray}
 (in the equations (6) from \cite{fw_old} only terms
  corresponding to considered interaction are kept).
 The solution is found by elementary methods:
 \begin{eqnarray}
   &&x^{+}(u)=A(f(u)+g(u)),\quad y^{+}(u)=A(f(u)-g(u)),\quad 
   x^{-}(u)=y^{-}(u)=Ah(u),\nn\\
   &&f(u)={{1+v^{2}}\over{1-v^{2}}}\cdot{{u}\over{1-u^{2}}}
   +\half\ln{{1+u}\over{1-u}},\quad
   g(u)={{2v}\over{1-v^{2}}}\left({{1}\over{1-u^{2}}}
   +{{v^{2}}\over{1-v^{2}}}\right),\label{prc1}\\
   &&h(u)={{u}\over{1-u^{2}}}+\half\ln{{1+u}\over{1-u}},\quad
   A={{(1-v^{2})^{5/2}}\over{4v^{3}}},\quad -1<u<1.\nn
 \end{eqnarray}
 The trajectories are symmetric {\it w.r.t.} frame origin
 (are conserved under the transformation
  $(x,y)^{\pm}\to-(y,x)^{\pm}\ \Leftrightarrow\ u\to-u$).
 Solution is given for CMF,
 $v$ is asymptotic velocity of the particles in CMF.
 General solution can be obtained from (\ref{prc1}),
 applying arbitrary Poincare transformations:
 \begin{eqnarray}
   &&(x,y)^{+}\to(x,y)^{+}\cdot C+D^{+},\quad 
   (x,y)^{-}\to(x,y)^{-}/C+D^{-}. \label{prc2}
 \end{eqnarray}
 The solutions (\ref{prc1}) have the following important feature:
 for $v\to1$ in zero velocity (laboratory) frame,
 depicted in \fref{proc}b,
 the trajectories have polygonal shape.
 Interaction localizes in a vicinity of fractures $A,A'$.
 Sewing this solution and $P$-reflected one together in points $B,B'$,
 a limiting Hill's solution for the problem
 with interaction of both types is obtained,
 see \fref{proc}c.

\vspace{5mm}\noindent
\ortograph{2.}
 Similarly one can obtain analytical description of limiting trajectories,
 presented in \fref{g5}.

 \vspace{3mm}
 \begin{figure}\label{lam}
a)~~~\epsfysize=3.5cm\epsfxsize=3.5cm\epsffile{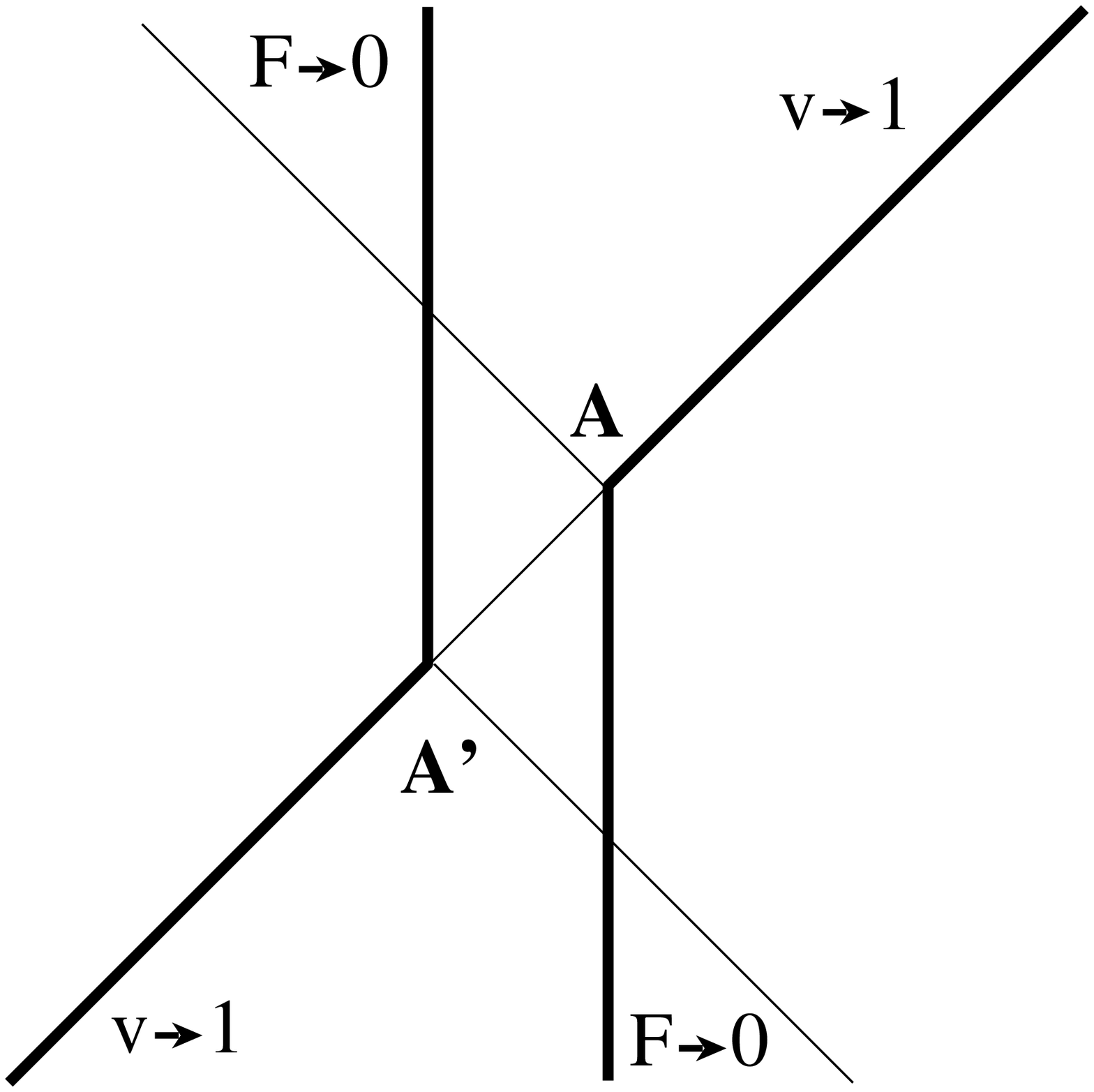}
~~b)~~~\epsfysize=3.5cm\epsfxsize=3.5cm\epsffile{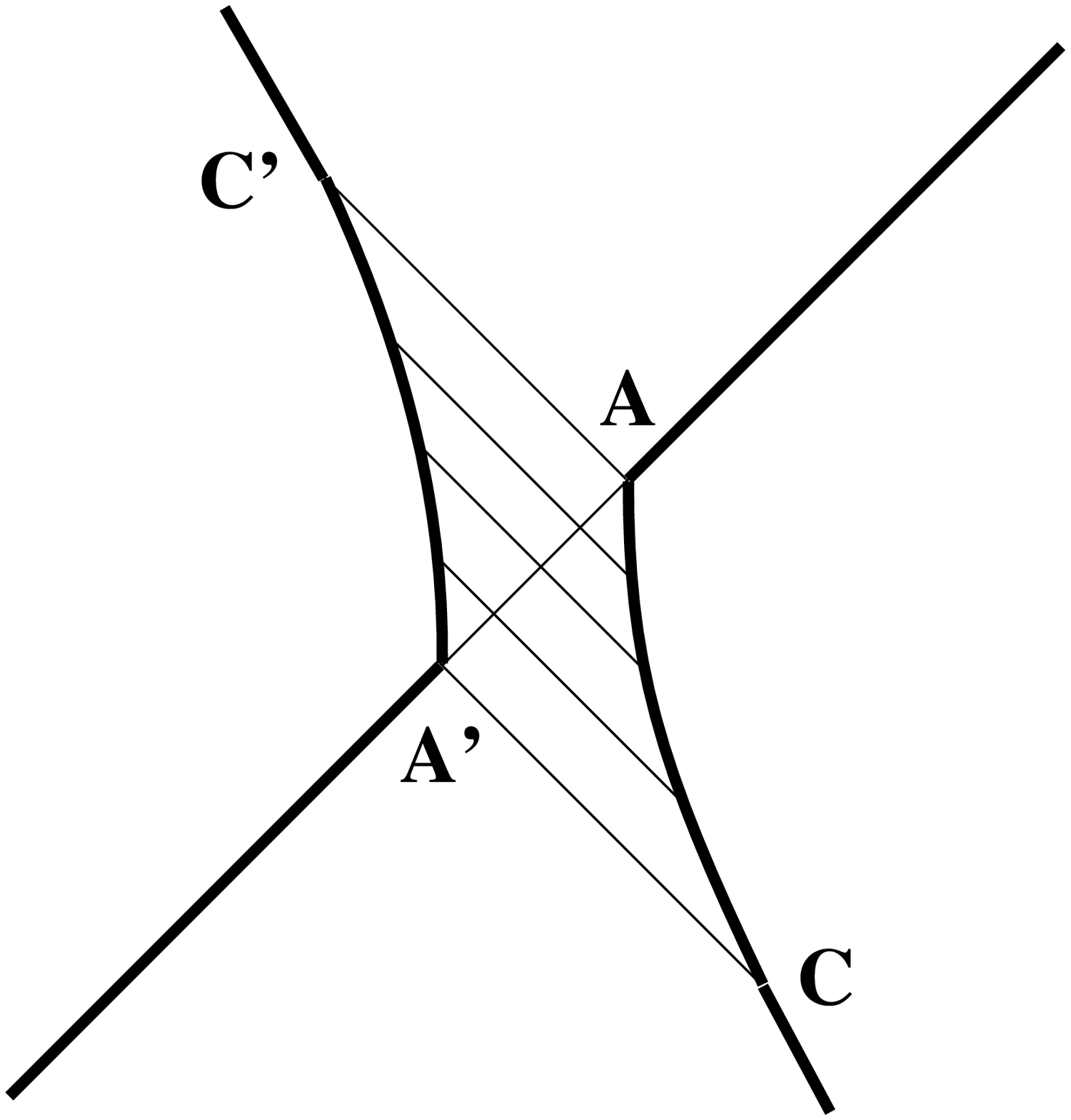}
~~c)~~~\epsfysize=3.5cm\epsfxsize=3.5cm\epsffile{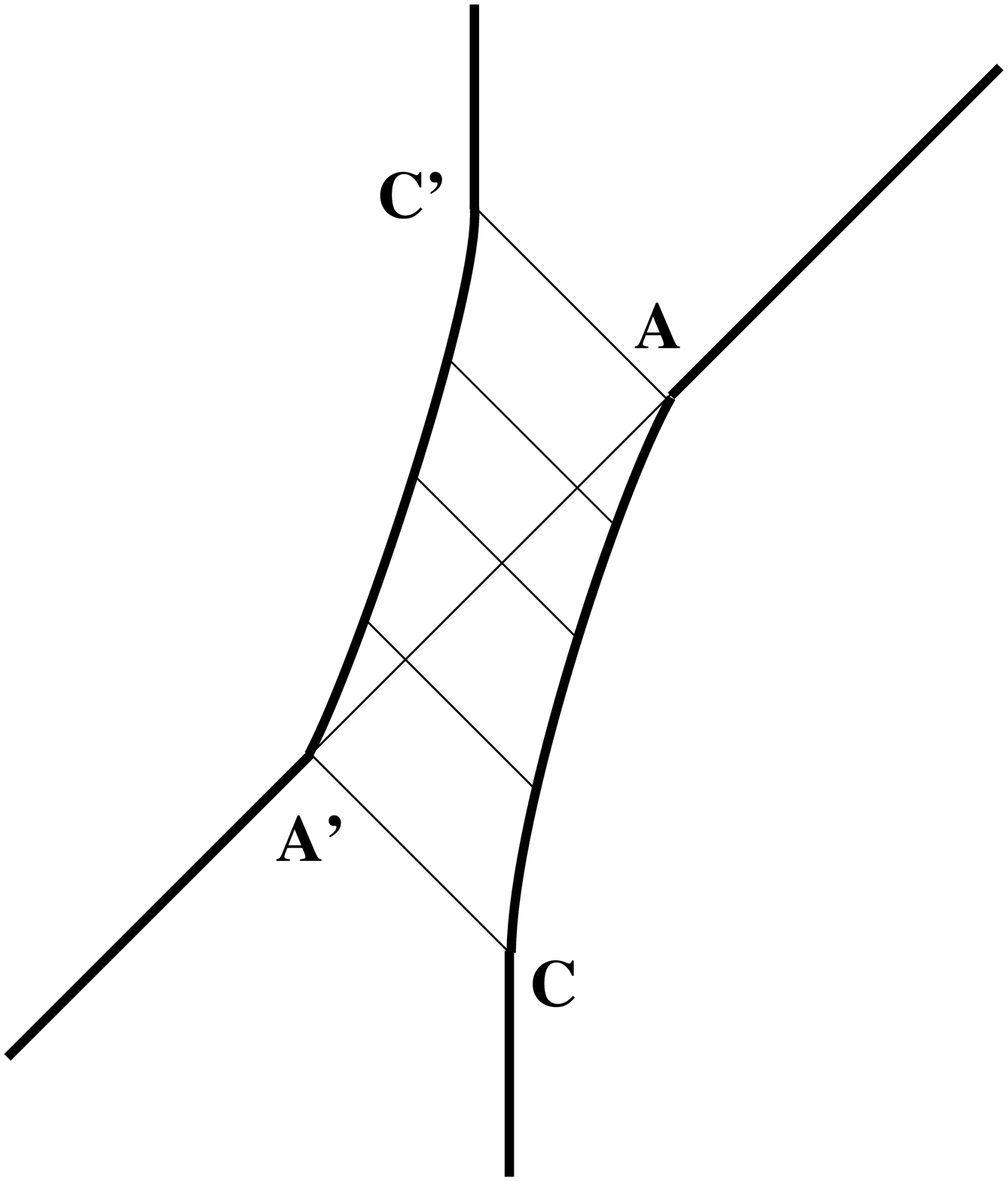}

   \vspace{0.3cm}
   \fignum 
 \end{figure}

 \vspace{3mm}
 Let's consider extremal trajectories for interaction of the 1-st type
 (\fref{lam}a) and take into account contribution of interaction of the
2-nd type.
 The trajectories regions, where particles move at light velocity,
 give no contribution to the interaction of the 2-nd type,
 because this interaction is suppressed by the factor $(1-v)/(1+v)\to0$
 for $v\to1$ (see (6) in \cite{fw_old}).
 In the internal intervals $CA,A'C'$ the interaction of the 2-nd type 
does not vanish,
 these trajectory regions have the shape, described by eq.(\ref{prc1}),
 transformed by $P$-reflection $(x,y)^{\pm}\to(y,x)^{\mp}$
 and transformations (\ref{prc2}).
 Outer intervals are straight and are sewed with internal ones
 smoothly in points $C,C'$, while in points $A,A'$ there are breaks
 of slope, of the same amplitudes as those in \fref{lam}a.

\begin{figure}\label{g6}
~\epsfysize=6.5cm\epsfxsize=6.5cm\epsffile{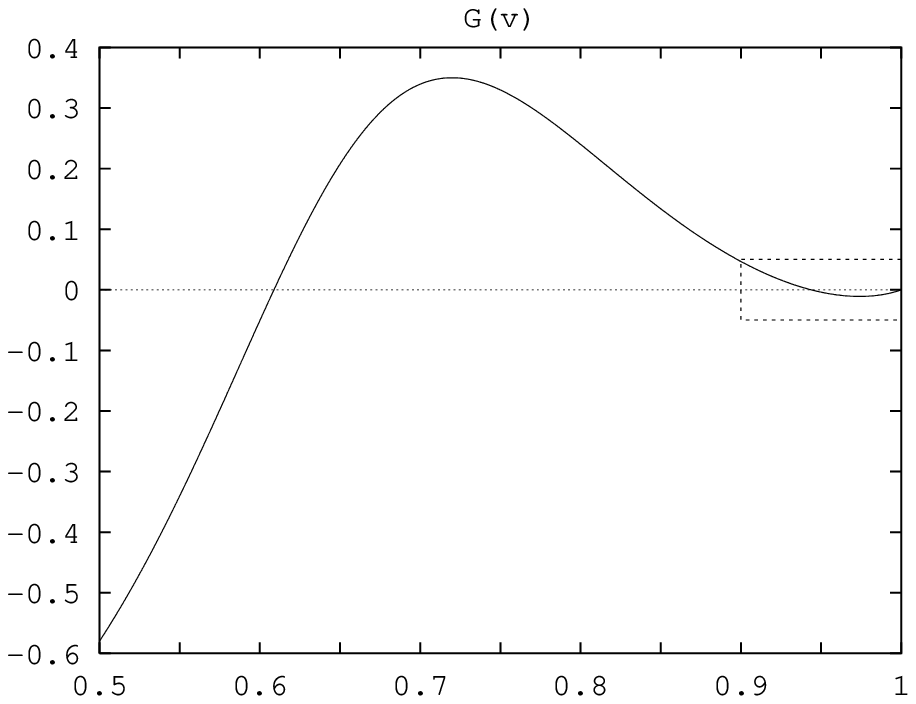} 
~\epsfysize=6.5cm\epsfxsize=6.5cm\epsffile{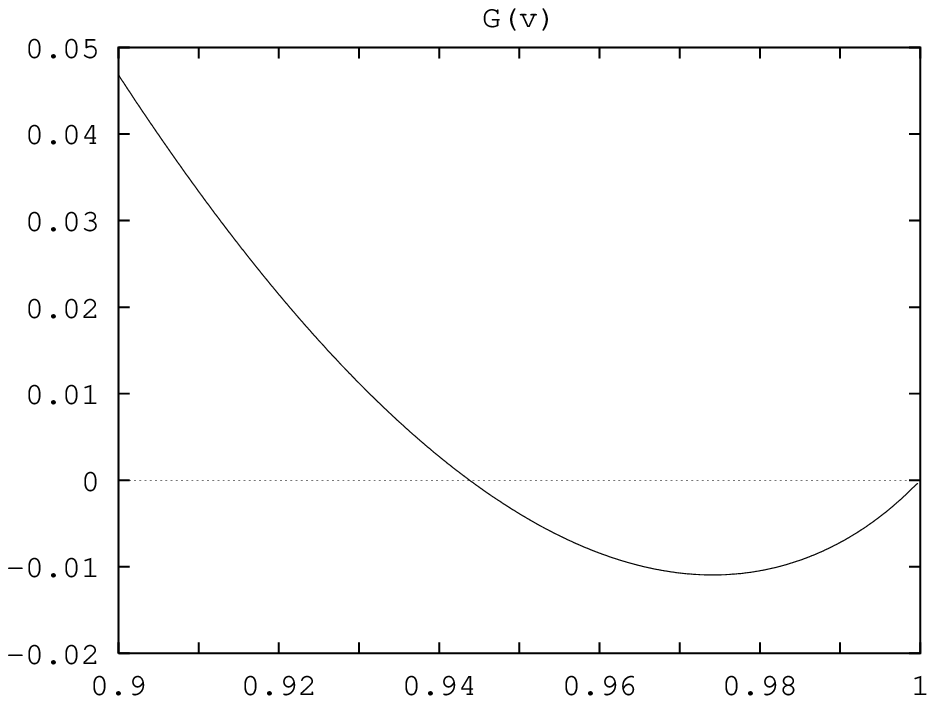}

   \vspace{0.3cm}
   \fignum 
 \end{figure}

 Thus, on internal intervals the solution has the form
 \begin{eqnarray}
   &&x^{-}=A(f-g)/C,\quad y^{-}=A(f+g)/C,\quad
   x^{+}=y^{+}=AhC.\label{prc3} 
 \end{eqnarray}
 (We have excluded translations $D^{\pm}$ from transformations (\ref{prc2})
  and further consider solutions symmetric {\it w.r.t.} frame origin.
  A more thorough analysis shows
  that there are no solutions with $D^{\pm}\neq0$).
 Requirements of sewing with straight trajectory regions
 have the form:
 \begin{eqnarray}
   &&x^{+}(u)=1,\quad x^{-}(u)=0,\quad {{dx^{+}}\over{dx^{-}}}(u)=1\quad
   \mbox{(point $A$ in \fref{lam}b)}\quad\Leftrightarrow\nn\\
   &&C=1/A(v)h(u),\ f(v,u)-g(v,u)=0,\
   G={{1-v^{2}}\over{(1-vu)^{2}A(v)^{2}h(u)^{2}}}-1=0.\label{prc4}
 \end{eqnarray}
 The function $x^{-}(u)$ in the interval $-1<u<1$
 monotonically increases from $-\infty$ to $+\infty$,
 thus the second equation in (\ref{prc4}) can be unambiguously resolved
 for $u$ (with $v$ fixed).
 Finding this solution numerically and substituting it to the third equation,
 we plot $G(v)$ function graph, see \fref{g6}.

 \vspace{3mm}\noindent

 \noindent
 \begin{figure}\label{g7}

\mar\epsfysize=3.8cm\epsfxsize=4cm\epsffile{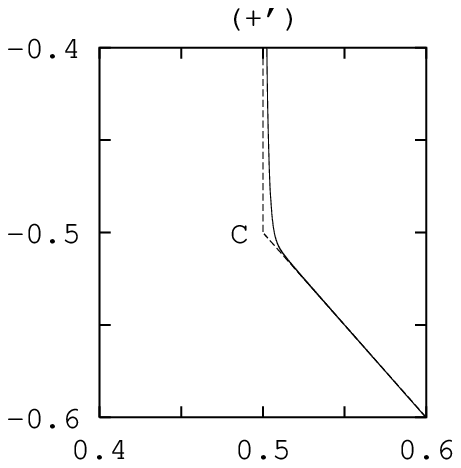}
\mar\epsfysize=3.8cm\epsfxsize=4cm\epsffile{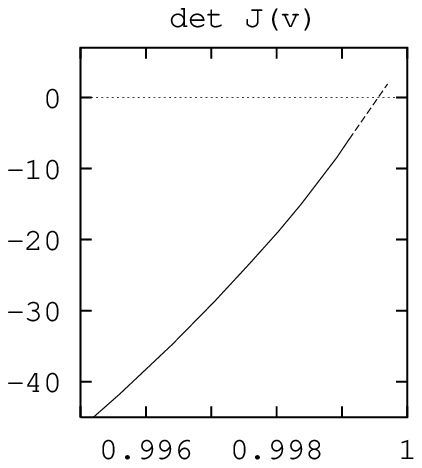}
   \quad\fignum 
 \end{figure}

 \vspace{3mm}\noindent
 The equation $G(v)=0$ has the following solutions\relax
 \footnote{Equations (\ref{prc4}) are satisfied with $10^{-15}$ accuracy
    on solution 2 and $10^{-14}$ accuracy on~3.}:

 \noindent
 1) $v\to1$ corresponds to Hill's solution.

 \noindent
 2) $v=0.6087622455403174\ (u=0.867439322472601,\ C=0.5948623884508324)$
 corresponds to limit of solution $(+)$.
 In order to transform this solution to laboratory reference frame \fref{lam}c,
 let's define the coefficient
 \begin{eqnarray}
   &&\tilde C=\left({{dx^{+}}\over{dx^{-}}}(\tilde u)\right)^{-1/2},\quad
   \mbox{where}\ x^{+}(\tilde u)=-1\ 
   \mbox{corresponds to point $C$ on \fref{lam}b.}\nn
 \end{eqnarray}
 Solving the last equation, we find
 $\tilde u=-0.8674393224726012, \ \tilde C=3.2378655431429$.
 Carrying out Lorentz transformations with coefficient $\tilde C$
 on the trajectories,
 we get limiting solution, depicted in top right of \fref{g5}.

 \noindent
 3) $v=0.9437848540619277\ (u=0.9998999343906364,\ C=0.1703458543931005)$,
    the solution is presented in \fref{g7}.
 In the reference system \fref{lam}b this solution is close to Hill's one.
 Small (1\%) discrepancies of these solutions in the interaction region
 are observed only in vicinities of points $C,C'$
 (also there is a small discrepancy of the solutions
  far from the interaction region,
  caused by difference of velocities $v_{x}$ for these solutions).
 In the laboratory reference system \fref{lam}c
 non-trivial limiting shape establishes for this solution as well
 (correspondent coefficient $\tilde C=34.5179309569148$).
 For transition to reference systems, where $v_{x}\to1$,
 coefficient $\tilde C\to0$ is required.
 This, particularly, means that in CMF at $v\to1$
 this solution collapses to the frame origin.

 {\it Our hypothesis}:
 for large $v$ the solution (0) undergoes one more bifurcation.
 As a result 3 solutions appear,
 one of them tends to Hill's solution at $v\to1$,
 while the others to new found solution and it's $P$-adjoint.
 The form of $\det J(v)$ dependence (see \fref{g7})
 conforms to existence of root near $v=0.9995$.
  The presented techniques can not continue the solution
to velocities $v>0.999$, due to computational difficulties 
described in \cite{fw_prog}.

\section{Proposal for further investigation}
 The main result of our work consists in the discovery of a fact,
 that the system considered has the same number of freedom degrees
 as it's non-relativistic analog
 (a system of two particles on a line,
  interacting via potential forces).

 In both cases the shape of the trajectories on a plane $(x,t)$
 depends only on particles asymptotic velocities
 (in our problem is defined by it uniquely for $v<0.937$
  and three-way-ly for $v>0.937$).
 Phase space of Hamiltonian theory in the studied problem
 is essentially 4-dimensional, as well as in non-relativistic case.
 The phase space structure (both topological, and symplectic),
 however, for these cases considerably differs.
 The scheme for construction of reduced phase space for the considered problem
 looks like the following.

 The set of constraints (including $H=0$ and sewing requirements)
 extracts 5-dimensional surface in extended phase space $(x,p)_{n}^{\mu}$.
 This surface looks like
 ${\cal S}_{5}=\Gamma_{1}\times{\cal P}_{3}\times\Tau_{1},$
 where $\Gamma_{1}=X(\mu)$ --- the found branchy curve\relax
 \footnote{see {\tt http://viswiz.gmd.de/\~{}nikitin/fw\_{}data/node1.html}},
 ${\cal P}_{3}$ --- Poincare group,
 $\Tau_{1}$ --- group of shifts along the trajectories.
 Reduction of symplectic form $dp_{n}^{\mu}\wedge dx_{n}^{\mu}$
 from extended phase space onto this surface
 follows to symplectic form, degenerate along the $\Tau_{1}$
 (solutions shifts are generated by Hamiltonian $H$,
  included into the full set of constraints).
 Factorization ${\cal S}_{5}$ by $\Tau_{1}$
 (or imposition of gauges like $\tau\sim x_{1}^{+}$)
 gives 4-dimensional phase space
 ${\cal S}_{4}=\Gamma_{1}\times{\cal P}_{3}$,
 where the reduced symplectic form defines Poisson brackets.

 \vspace{3mm}
 \begin{figure}\label{f12}
   \begin{center}
~\epsfysize=2.4cm\epsfxsize=6cm\epsffile{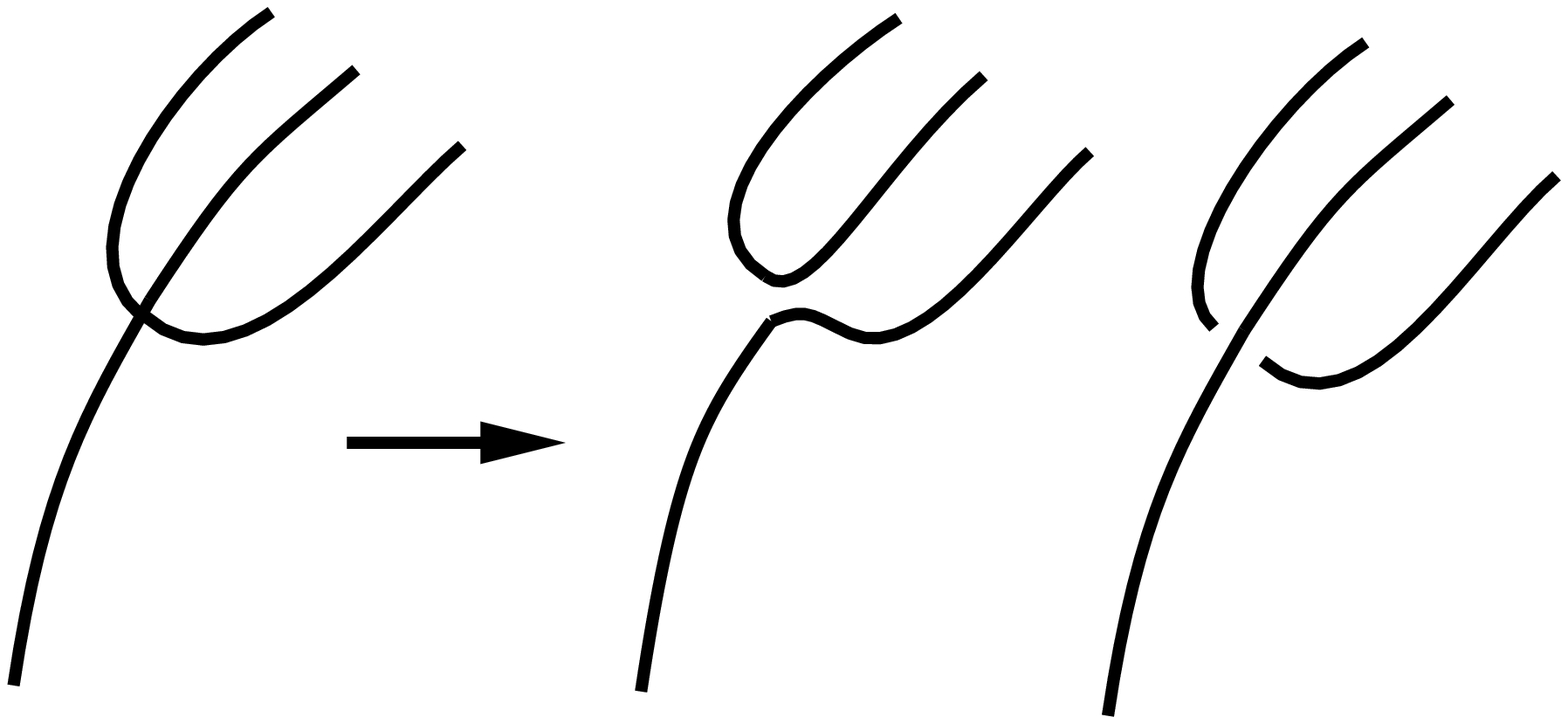}
      \quad\fignum 
   \end{center}
 \end{figure}

 \vspace{3mm}
 In further investigations of the given problem
 it's interesting to carry out:

 \noindent
 1.
 Calculation of Poisson brackets for independent variables
 and construction of canonical basis
 in the obtained Hamiltonian mechanics.

 \noindent
 2.
 A study of non-symmetrical problem $m_{1}\neq m_{2}$.
 The fact that the branches $X^{(0,\pm)}(\mu)$ of found solution 
 are sewed together in one point of phase space
 is related to the symmetries of the problem.
 For $m_{1}\neq m_{2}$ in a vicinity of critical point
 recombination and formation of disjoint branches of solution are 
possible (see \fref{f12}).

 \noindent
3.
 Calculation of second variation of the action 
 to investigate it's extrema.
 Necessity of such investigation consist in the following.
 Let's consider smooth function $f:\ \R^{n}\to\R$,
 which has $k>1$ extrema in $\R^{n}$.
 In this case all extrema can not be minima,
 there should be extrema of other type among them (maxima or saddles).
 One should expect that not all of found solutions of the problem 
 correspond to minimum of action,
 there should be extrema of other types as well.

 \noindent
 4.
 Investigation of system responses to external perturbations
 and analysis of causal properties of the system.
 As advanced interactions can lead to violation of causality principle,
 it's interesting to answer the following question:
 can an observer related to one of the trajectories
 detect the external influence on it before it takes place?
 In the case of causality violation
 one should estimate the magnitude of the effect.

 \baselineskip=\normalbaselineskip

\end{document}